\documentclass[lettersize,journal]{IEEEtran}
\usepackage{amsmath,amsfonts}
\usepackage{algorithmic}
\usepackage{algorithm}
\usepackage{array}
\usepackage{textcomp}
\usepackage{stfloats}
\usepackage{url}
\usepackage{verbatim}
\usepackage{graphicx}
\usepackage{cite}
\usepackage{multirow}
\usepackage{svg}
\usepackage{quoting}
\usepackage[dvipsnames]{xcolor}

\usepackage{subcaption}
\usepackage{float}
\usepackage{booktabs}

\usepackage{hyperref}

\begin{document}

\quotingsetup{vskip=3pt}

\newcommand{\interfaceName}[1]{\textit{#1}}
\newcommand{\defn}[1]{\textbf{#1}} 

\newcommand{\SuppFigMCVDetails}[0]{Supp.~Fig.~S2} 
\newcommand{\SuppSecJoins}[0]{Supp.~Tab.~S1}
\newcommand{\SuppSecWorkflows}[0]{Supp.~Sec.~S3}
\newcommand{\SuppSecDeployment}[0]{Supp.~Sec.~S2}
\newcommand{\SuppSecUsability}[0]{Supp.~Sec.~S2.2}
\newcommand{\SuppFigColDisplay}[0]{Supp.~Fig.~S1}

\newcommand{\MegaFig}[1]{Fig.~4~(#1)}
\newcommand{\MegaFigOpsTab}[0]{\MegaFig{a}} 
\newcommand{\MegaFigColsTab}[0]{\MegaFig{b}} 
\newcommand{\MegaFigCTS}[0]{\MegaFig{c}} 
\newcommand{\MegaFigTableSchemaCompressed}[0]{\MegaFig{d}} 
\newcommand{\MegaFigTableSchemaCondensed}[0]{\MegaFig{e}} 
\newcommand{\MegaFigPackSchema}[0]{\MegaFig{f}} 
\newcommand{\MegaFigTableRows}[0]{\MegaFig{g}} 
\newcommand{\MegaFigDQTab}[0]{\MegaFig{h}} 
\newcommand{\MegaFigMetadataTab}[0]{\MegaFig{i}} 
\newcommand{\MegaFigDistributionTab}[0]{\MegaFig{j}} 
\newcommand{\MegaFigValueCounts}[0]{\MegaFig{k}} 

\newcommand{\InterfaceFigRef}[1]{Fig.~\ref{fig:interface}~(#1)}
\newcommand{\DataInventoryFig}[0]{\InterfaceFigRef{a}}
\newcommand{\SchemaFig}[0]{\InterfaceFigRef{b}}
\newcommand{\TableRowsFig}[0]{\InterfaceFigRef{c}}
\newcommand{\CompositeSchemaFig}[0]{\InterfaceFigRef{d}}
\newcommand{\ColumnComparisonFig}[0]{\InterfaceFigRef{e}}

\title{OpenRoundup: Multi-Table Data Wrangling Through Interactive Visualization}

\author{Stephen Kasica, 
    Charles Berret, 
    Tamara Munzner
    \IEEEcompsocitemizethanks{
        \IEEEcompsocthanksitem
        Stephen Kasica and Tamara Munzner are with the University of British Columbia, Department of Computer Science. E-mail: kasica@alumni.ubc.ca, tmm@cs.ubc.ca.
        \IEEEcompsocthanksitem
        Charles Berret is with Enigma Technologies, Inc. E-mail: charles.berret@enigma.com.}
    \thanks{Manuscript received XXX; revised XXX.}}

\markboth{TVCG Journal Submission, Jun~2026}%
{Kasica \MakeLowercase{\textit{et al.}}: OpenRoundup}


\maketitle

\begin{abstract}
Data journalists routinely integrate records across multiple independently published sources to support accountability reporting, yet no existing interactive wrangling tool treats the collection of tables — rather than the single table — as its primary unit of work. We present OpenRoundup, an open-source, browser-based system that enables data journalists to consolidate multiple tables into a single analysis-ready output without writing code. The interface comprises five coordinated panels that implement a schema-first, values-on-demand paradigm with live schema previews, ambient data quality alerts, and a recursive treemap visualization of the evolving operation tree. A client-only architecture powered by DuckDB-WASM runs in the browser, providing strong data privacy guarantees suited to sensitive journalism data. The system introduces two conceptual contributions: eager table consolidation, in which a composite table is assembled early in the wrangling phase via  interactive, incremental assembly of multiple source tables; and a declarative vocabulary for table consolidation consisting of two operations, Stack and Pack. We evaluate the system through a replication study in which the authors reproduce 17 published journalist programming workflows using only the interface, and a deployment study with four professional data journalists. The replication study demonstrates expressive coverage of real-world consolidation tasks. The deployment study confirms utility for practitioners who understand joins conceptually but lack the programming skills to execute them, and surfaces an unanticipated secondary value for data journalism education.
\end{abstract}


\section{Introduction}
\label{sec:introduction}

In \defn{data journalism} --- the practice of combining data collection, analysis, and computational methods with traditional reporting --- the stories that matter most are rarely contained in a single dataset. A central challenge is \defn{table consolidation}: the problem of integrating records across multiple heterogeneous sources into a unified, analysis-ready structure. Accountability reporting, namely investigations into institutions, governments, and systems on behalf of the public interest, has long depended on this kind of cross-source record linkage~\cite{cohen-computational-2011}. Examples include matching school bus drivers to criminal records, recidivism risk scores to criminal histories, sugar mill owners to offshore shell companies~\cite{meyer-precision-2002, bounegru-data-2021}.

A crucial support for this practice
is \defn{data wrangling}: the iterative process of exploring, cleaning, and transforming data into a usable form~\cite{kandel-research-2011}.

However, existing tools provide poor support for journalists engaged in table consolidation, because code-based wrangling tools are inaccessible to the majority of journalists. 
Data work has long been regarded as a specialist skill requiring extensive training~\cite{rogers-data-2017}; 
in practice, fewer than one in three data journalists use programming for their work and over half consider themselves novices or lacking skills entirely in data wrangling~\cite{ejc-2021}. Only few data journalists receive formal education in technical fields~\cite{heravi-lorenz-2020}, and this non-technical background is reflected in tool adoption: spreadsheet tools far outpace code-based alternatives, with spreadsheet applications used by the majority of data journalists~\cite{ejc-2021}. 

While interactive wrangling interfaces have lowered the barrier to data preparation for non-programmer audiences, they treat the single table as the unit of work. They support within-table wrangling tasks such as reshaping, cleaning, and type conversion, but provide limited support for cross-table integration. Multi-table operations are pervasive in journalism workflows, yet remain absent from the conceptual frameworks of the interactive wrangling literature~\cite{kasica-table-2021}.

Conversely, extract, transform, and load (ETL) systems, pipelines that move and reshape data from source systems into a central repository for storage and analysis, do support multi-table integration, but they are designed for different users and contexts. First, they assume programming literacy and data-modeling expertise. Moreover, these systems are built for enterprise data warehouses where schemas are standardized within an organization~\cite{furche-data-2016}. By \defn{schema}, we mean the structure of a table: a tuple of column metadata, including names, declared data types, and semantics, and the row values each column contains. 

While journalists sometimes maintain their own databases, often due to missing data~\cite{dignazio-data-2020}, the return on investment for setting up a persistent pipeline is low for one-off stories where the initial cost cannot be amortized across future use. Thus, the common case for journalists is to use data outside the control of their own organization, drawn from public, third-party sources. In these cases, the target schema is unknown and must be discovered through the wrangling process. 

Also, journalists may need to conduct pre-publication analysis locally, 
for sensitive data under embargo until publication.

The combination of these constraints --- the non-technical background of the user group, the schema heterogeneity of public sources, the transient nature of journalistic data work, and need for local processing --- define a research-to-reporting gap~\cite{stray_making_nlp_2017} that existing tools do not address.

We present \textbf{OpenRoundup}, an open-source, browser-based application that enables data journalists to consolidate multiple tables without writing code. Users upload a collection of source tables, incrementally assemble them into a unified structure through a direct manipulation interface, and export a single analysis-ready table for downstream use. Throughout this process, OpenRoundup interactively surfaces potential schema conflicts and data quality issues, making the consequences of integration decisions visible before they propagate.

The primary contribution of this paper is the design, architecture, and implementation of the OpenRoundup system. The system is realized across five coordinated panels that provide live schema previews, ambient data quality alerts, and a recursive treemap visualization of the evolving operation tree. The secondary contributions of this paper are twofold. First, a new conceptual framing that distinguishes this work within the data integration problem: \defn{eager table consolidation}, in which a composite table is assembled early in the wrangling phase via interactive, incremental assembly of multiple source tables. Second, an expressive and minimal vocabulary for assembling tables consisting of two operations, \textit{stack} and \textit{pack}, which are amenable to visual representations. We evaluate OpenRoundup through a replication study in which the authors replicate 17 journalistic programming workflows using OpenRoundup, and a deployment study with four professional data journalists.

\section{Related Work}
\label{sec:related-work}

We situate OpenRoundup within three bodies of prior work: interactive data wrangling systems that have lowered the barrier to data transformation for non-programmer audiences, domain-specific tools built for journalism data workflows, and the broader data integration literature addressing multi-source schema alignment. Each tradition addresses part of the problem space that OpenRoundup occupies.

\subsection{Interactive Data Wrangling Systems}
\label{sec:wrangling-systems}

A lineage of systems has lowered the barrier to data transformation through interactive direct manipulation. Potter's Wheel~\cite{raman-potters-2001} first established a closed feedback loop of transform, profile, and repeat. Extending this approach with mixed-initiative inference, Wrangler~\cite{kandel-wrangler-2011}~\footnote{later commercialized as Trifacta and acquired by Alteryx in 2022} infers and rank-orders applicable transformations as the user manipulates data. Guo et al.~\cite{guo-proactive-2011} further enriched inference by treating the input data format of downstream analysis tools as an additional ranking signal. Rather than transformation, Profiler~\cite{kandel-profiler-2012} shifted attention to assessment, investigating automated methods for surfacing data quality issues. Rigel~\cite{chen-rigel-2022} took a declarative approach: instead of specifying procedural transformation sequences, users describe the desired output structure. Most recently, Buckaroo~\cite{warner-buckaroo-2025} identifies anomalous subgroups in tabular data, suggests repair operations, and visualizes the downstream impact of each fix in real time. Across these systems, a design commitment persists: intermediate state must remain visible during transformation~\cite{kandel-research-2011}. This principle extends to code-based settings as well: Xavier~\cite{zhou-xavier-2025} and Unravel~\cite{shrestha-unravel-2021} demonstrate that exposing intermediate results is critical when wrangling is expressed programmatically. However, all of these approaches focus on a single table. 

OpenRoundup carries this commitment into the multi-table context: all operations provide live schema previews, row-count updates, and alert feedback, ensuring the evolving composite structure remains visible throughout consolidation. OpenRoundup departs from the Wrangler lineage's procedural, transformation-log model in favor of a declarative interaction model, following Rigel~\cite{chen-rigel-2022}, and applies it to a collection of tables rather than to an individual one. In contrast to these transformation-focused systems, TACO~\cite{niederer-taco-2018} addresses a complementary problem: visualizing structural and content differences across multiple versions of the \textit{same} table over time, a concern orthogonal to OpenRoundup's focus on consolidating \textit{across} independently-structured sources.

\subsection{Data Preparation Tools for Journalists}
\label{sec:journalism-tools}

While spreadsheet applications dominate data work in newsrooms, particularly Microsoft Excel~\cite{mulvad-effective-2018}, journalists also use a variety of other tools to a lesser extent. OpenRefine, formerly known as Google Refine and Freebase Gridworks, is a popular open-source tool for interactive data cleaning and transformation with a small but consistent base of journalist users~\cite{magdinier-openrefine-2024}. Tableau Prep Builder\footnote{\url{www.tableau.com/products/prep}}, a commercial data preparation tool from Tableau, supports multi-source cleaning through an imperative pipeline model and is licensed for free to journalists who are members of The Investigative Reporters \& Editors (IRE) organization.

A small ecosystem of open-source data manipulation software written for journalists by journalists has also emerged. Tabula~\cite{tabula} extracts tables from PDFs, which often introduces errors that require data cleaning in downstream tools, such as OpenRoundup. \texttt{csvkit}~\cite{groskopf-csvkit-2012} provides command-line CSV transformations, and \texttt{agate}~\cite{groskopf-agate-2018} offers data transformation in Python designed specifically for journalism workflows. The Workbench\footnote{\url{github.com/CJWorkbench/cjworkbench}} project from the Columbia Journalism School provided an integrated interactive platform spanning scraping, cleaning, analysis, and visualization. OpenRoundup belongs to this community-oriented tradition, as two of its authors also have newsroom experience, and addresses the table consolidation gap that all of these tools leave open.

\subsection{Data Integration Systems}
\label{sec:integration-systems}

OpenRoundup is related to the broader data integration literature~\cite{halevy-teenage-2006,stonebraker_data_2018}, but differs fundamentally from the assumptions and design patterns of this tradition. Classical data integration pipelines proceed through three sequential phases: schema mapping, duplicate detection, and data fusion~\cite{bleiholder_data_2009}. OpenRoundup deliberately addresses only the first phase. Journalism's one-off, ad-hoc workflows have no downstream warehouse or persistent pipeline to serve, making the full pipeline model both architecturally over-specified and practically inaccessible to non-programmer users. We review three relevant strands of work.

\paragraph{Schema mapping} Identifying semantically corresponding columns across multiple tables is central to both SQL joins and unions. Despite advances in automating this task, a human in the loop remains necessary~\cite{rahm-survey-2001}. At enterprise scale, that human involvement becomes the limiting factor~\cite{stonebraker_data_2018}. Clio~\cite{fagin-clio-2009} semi-automatically generates declarative mappings between relational and XML schemas, compiling them to SQL, XQuery, or XSLT. Chiticariu et al.~\cite{chiticariu-interactive-2008} extend this line of work by enumerating multiple candidate integrated schemas rather than producing a single result. SchemaMapper~\cite{robertson-schemamapper-2005} uses highlight propagation and auto-scroll to make large enterprise XML-to-XML maps manageable, addressing scalability for experts rather than accessibility for novices. Recognizing the need for lightweight schema mapping among non-expert users, Qian et al.~\cite{qian-sample-driven-2012} propose a sample-driven system where users type example target instances and the system infers join paths. 

The interaction models in these systems share a critical assumption: users have a preconceived notion of a target schema, which is not true for the table consolidation context. In contrast, OpenRoundup mediates this emergent schema through live previews, alerts, and a visual operation tree. This exploratory requirement, of discovering a schema rather than mapping between two known ones, distinguishes table consolidation as a wrangling task from schema mapping. 

Data lake architectures~\cite{fang-managing-2015} partially relax the schema requirement through a schema-on-read model, but they retain the same assumptions about technical expertise and persistent organizational infrastructure. Moreover, they are server-side systems that are incompatible with the data privacy demands of pre-publication journalism. 

\paragraph{In-situ integration} Several systems attempt to avoid the overhead of data preparation via integration \textit{in-situ}, within visual analytics tools. Coscia et al.~\cite{coscia-preliminary-2024} found that in-situ approaches allowed users to spend less time on data integration compared to traditional \textit{ex-situ} workflows. Morton et al.~\cite{morton-dynamic-12} approached this problem in the commercial space with \textit{data blending} in Tableau via an interface for multi-source integration with automatic key detection and on-the-fly aggregation. 

While in-situ integration works well for well-structured business data with consistent column naming, it fails when column names diverge, join keys do not exist, or tables have no established linkage, all typical conditions for datasets used in journalism. OpenRoundup adopts an ex-situ model that makes integration the primary activity, supporting complex join types and dirty data~\cite{kasica-dirty-2023}.

\paragraph{End-user integration} Augmentation tools~\cite{tuchinda-building-2008,lin-end-user-2009,ives-interactive-2009} demonstrate that non-experts can integrate data without programming, but they target web data extraction and data foraging, finding well-structured data via the Web of Linked Data~\cite{bizer-emerging-2009}. OpenRoundup's user already has the data and needs to consolidate it. Potluck~\cite{huynh-potluck-2007} is the closest in spirit to our system: casual users drag field tags to align heterogeneous RDF data from multiple sources while building visualizations, demonstrating the viability of direct-manipulation schema alignment for non-experts. Van Kleek et al.~\cite{van-kleek-carpe-2013} similarly target non-expert data mixing, using drag-and-drop co-reference resolution to reconcile terminologically heterogeneous records from web data feeds; their finding that any integration approach requiring sources to be specified \textit{a priori} is unsuitable for exploratory workflows parallels our own design rationale. OpenRoundup extends these ideas to journalists' tabular data with explicit Stack/Pack semantics and no dependency on Semantic Web infrastructure. Knowledge graph systems~\cite{cashman-cava-2021,xia-ktabulator-2021, huynh-openrefine-2012} enable non-experts to augment tables using structured web data, but this kind of data reconnaissance~\cite{crisan-uncovering-19} is not the goal of the OpenRoundup system, which assumes users have already collected all necessary source data.

\section{Conceptual Framing}
\label{sec:framing}

We introduce the term \defn{table consolidation} to characterize the distinct integration challenge journalists face when combining data from independently published sources with no predetermined target schema. We propose \defn{eager consolidation} as a strategy for when consolidation should occur in a wrangling pipeline, realized through a \defn{snowball approach} to interactive assembly, and formalize the process using two declarative operations: \textbf{Stack} and \textbf{Pack}.

\subsection{Eager Table Consolidation}
\label{sec:table-consolidation}

Data integration is a well-studied problem, but the existing literature models it in ways that do not match the conditions journalists work under. Classical data integration addresses the problem of combining data from heterogeneous sources into a unified representation suitable for analysis or visualization~\cite{hendler-data-2014}. It encompasses schema alignment, entity resolution, and value harmonization across datasets that may differ in structure, format, provenance, or granularity. The defining assumption of this tradition is that both source and target schemas are known in advance; integration is a transformation from one understood structure to another~\cite{rahm-survey-2001}. 

These assumptions do not hold in journalistic data wrangling for structural rather than incidental reasons, and we introduce \defn{table consolidation} to capture this distinct integration context. Data journalists work with ASCII and binary files drawn from independently published sources that were never designed to be linked~\cite{showkat-where-2021}. Inter-table relationships must be identified by the analyst rather than declared via foreign key constraints. Data quality cannot be assumed. Misspellings, format inconsistencies, and missing values due to manually entered data are endemic to the administrative and government sources journalists regularly encounter~\cite{kasica-dirty-2023}. Because source data is outside of the control of a journalist's own organization, there is no pre-determined target schema to transform toward. The output schema in table consolidation is a product of the process, negotiated interactively as the user comes to understand the data, which corresponds to schema articulation in critical data sensemaking~\cite{berret-iceberg-2024}. Rather than a persistent pipeline or data warehouse, the goal of table consolidation is a single ad-hoc table for downstream analysis.

Every multi-table wrangling pipeline faces the structural choice of when consolidation occurs relative to other wrangling tasks. We distinguish two temporal approaches. In \defn{delayed consolidation}, individual tables are cleaned and transformed independently, before table consolidation. This approach maps onto the functional pipeline model common in scripting environments: a single cleaning function is mapped over a collection of source tables, keeping per-source transformation logic modular and the final consolidation step decoupled from upstream preparation.

We propose the alternative of \defn{eager consolidation}, in which a single composite table is assembled as early in the process as possible, exposing the unified schema to all subsequent wrangling. Only wrangling tasks that are strictly necessary such as table shaping and aggregation are performed upstream of consolidation, while all other tasks are deferred until the composite table is available. Surfacing the schema early allows users to identify structural mismatches, redundant columns, and join key compatibility before investing effort in downstream cleaning. This approach also supports workflows where a variety of data wrangling tools are used in combination, a common pattern in journalism workflows~\cite{kasica-dirty-2023}, and a trend in the broader data community through interchangeable data formats such as Apache Arrow\footnote{\url{https://arrow.apache.org/}}.

Eager consolidation, however, requires a mechanism suited to schema uncertainty: the user cannot specify a complete consolidation plan up front because the composite schema is itself discovered through the process. OpenRoundup therefore realizes eager consolidation through a \defn{snowball approach} to interactive table assembly, implemented as an incrementally constructed \defn{composite tree}. The user begins by combining two tables and iteratively merges additional tables into the growing composite, inspecting intermediate results at each step. The snowball metaphor captures the dynamic of a composite table that starts small and accumulates structure as it picks up additional source tables.

\begin{figure}[h]
  \centering
  \begin{subfigure}[t]{0.38\columnwidth}
    \includegraphics[width=\linewidth]{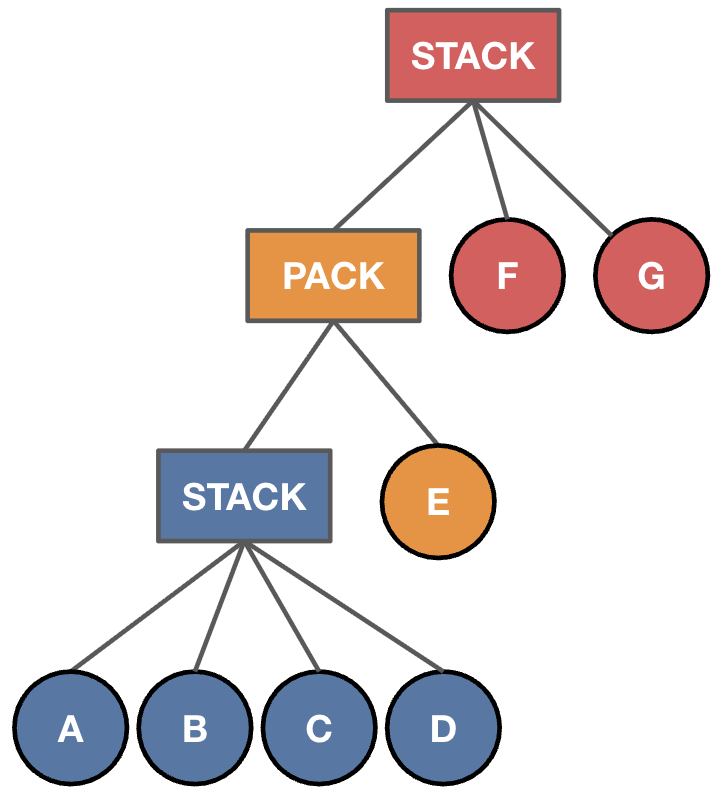}
    \caption{Operation tree, built from Pack and variadic Stack}
    \label{fig:operation-tree-compressed}
  \end{subfigure}
  \begin{subfigure}[t]{0.18\columnwidth}
    \includegraphics[width=\linewidth]{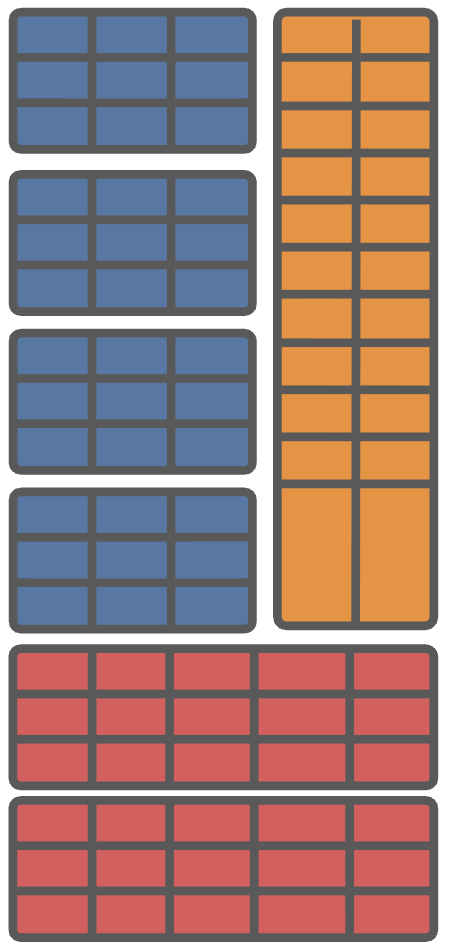}
    \caption{Compact composite table}
    \label{fig:operation-tree-tables}
  \end{subfigure}
  \begin{subfigure}[t]{0.39\columnwidth}
    \includegraphics[width=\linewidth]{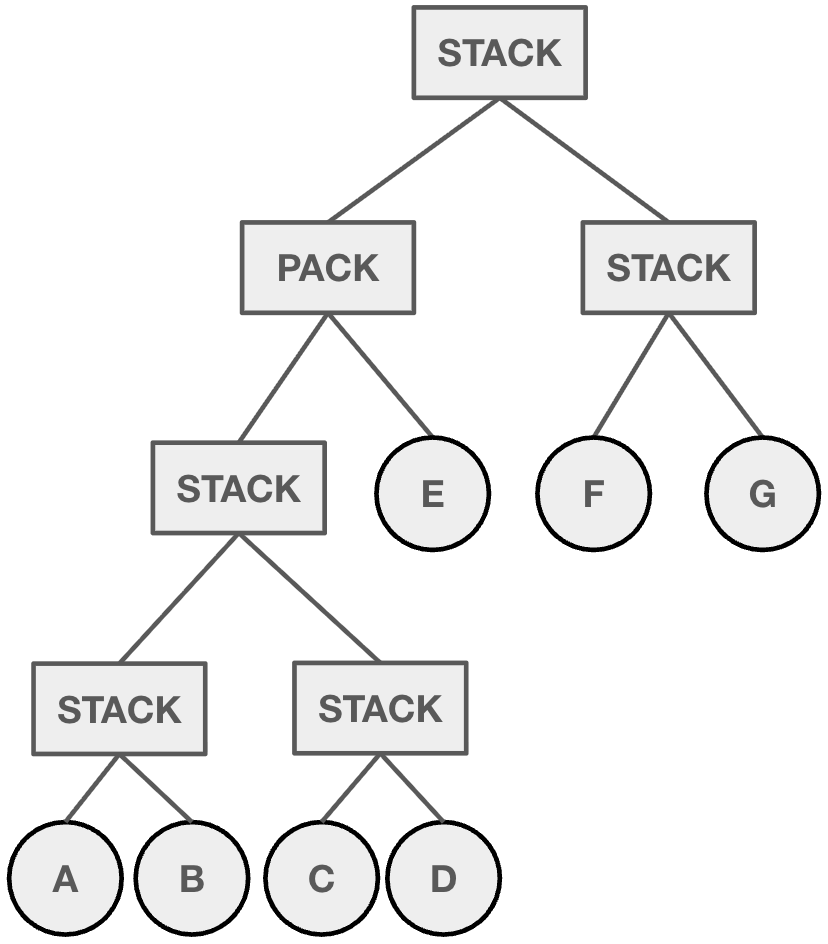}
    \caption{Naive binary design is taller than variadic version}
    \label{fig:operation-tree-naive}
  \end{subfigure}
  \caption{OpenRoundup supports interactive table assembly with a snowball approach.  (a)~The combination of Stack operations that append rows and Pack operations that combine horizontally by matching rows on shared columns yields an operation tree. (b) The resulting arrangement of source tables is shown as a compact composite table in OpenRoundup. (c)~A naïve binary design would require strictly $n - 1$ operations, where $n$ is the number of source tables to combine, in contrast to our variadic design with a flatter tree structure.}
  \label{fig:operation-tree}
\end{figure}

Each incremental step follows a four-stage interaction sequence: (S1) select the tables to combine, (S2) choose an operation, (S3) define the operation schema, and (S4) materialize the result. \defn{Materialization} executes all operations on the selected tables, producing an assembled table that can be inspected. This cycle supports the exploratory character of journalistic data work, where decisions about schema alignment are made empirically, in response to what the data reveals.

The output of the assembly process is an \defn{operation tree}, as shown in Fig.~\ref{fig:operation-tree-compressed}: a special case of a directed acyclic graph in which each node has in-degree at most one and intermediate results are not shared across branches. Operations are root and interior nodes, and source tables are one level higher than the leaves. Tables are composed of Columns, which are the third type of data that is surfaced in the OpenRoundup interface, and are the leaf nodes of the operation tree. The user produces this tree incrementally through interactions with the OpenRoundup interface, and a post-order depth-first traversal of the tree yields the equivalent relational algebra expression used to process tables. The resulting arrangement of source tables can be shown visually as a compact composite table, as in Fig.~\ref{fig:operation-tree-tables}.

\subsection{Pack and Stack: A Declarative Vocabulary for Table Consolidation}
\label{sec:pack-stack}

The snowball approach requires an operation vocabulary that is both expressive and minimal. Rather than exposing a general relational algebra, OpenRoundup abstracts the primary structural cases of journalistic consolidation into two declarative operations, shown in Fig.~\ref{fig:operation-tree}. The user specifies \emph{what} to combine rather than \emph{how} to execute it, keeping the interface interpretable to users without SQL or relational algebra knowledge.

The variadic \textbf{Stack} operation vertically combines two or more tables by appending rows, targeting tables with near-homogeneous schemas, analogous to SQL \texttt{UNION}. The binary \textbf{Pack} operation horizontally combines two tables by matching rows on a shared column, targeting tables with heterogeneous schemas that share a common key, analogous to SQL \texttt{JOIN}. Together, Stack and Pack cover the primary structural patterns observed in journalistic consolidation workflows: assembling data collected across time or region (Stack), and linking records across thematically distinct datasets (Pack)~\cite{kasica-table-2021}.

Because operations may be variadic, the composite tree remains shallow even for complex consolidations involving many tables. This design reduces tree depth and total node count while preserving full expressive power. A naïve binary-only design would require additional intermediate nodes to represent the same assembly, increasing tree depth without expressive gain, as shown in Fig.~\ref{fig:operation-tree-naive}.

\section{Interface Design}
\label{sec:interface}


\begin{figure*}[t]
  \centering
  \includegraphics[width=1\textwidth]{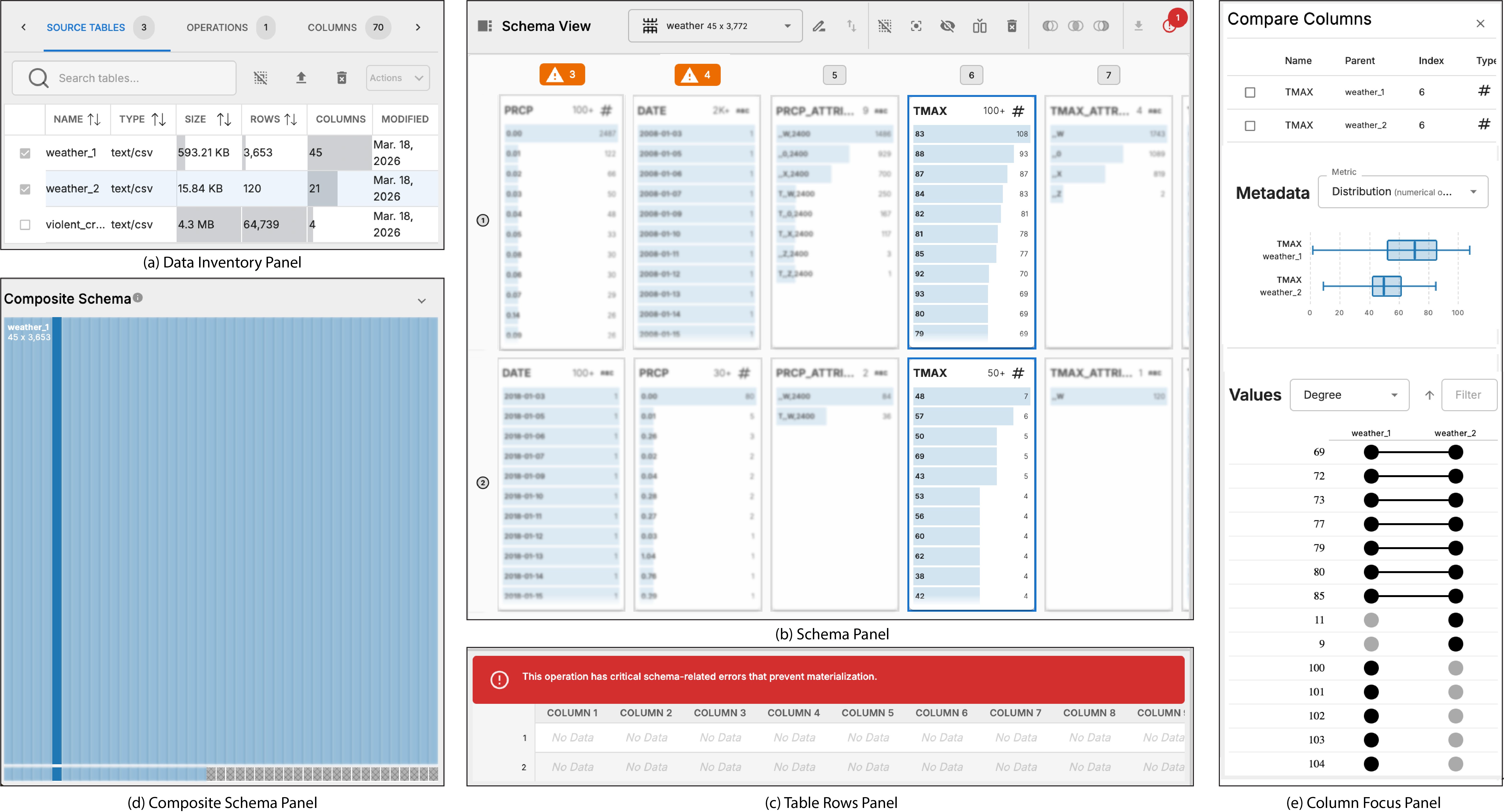}
  \caption{The OpenRoundup interface, showing an intermediate step in the Crime \& Heat usage example (Sec.~\ref{sec:crime-heat}). The OpenRoundup interface is organized into five panels: (a) the \interfaceName{Data Inventory Panel}, showing tabs for Source Tables, Columns, and Operations; (b) the central \interfaceName{Schema Panel} displaying column cards for the focused table or operation; (c) the \interfaceName{Table Rows Panel} below it showing a spreadsheet view of selected columns; (d) the \interfaceName{Composite Schema Panel} visualizing the operation tree as a treemap; and (e) the on-demand \interfaceName{Column Focus Panel} providing detailed analytics for selected columns.}
  \label{fig:interface}
\end{figure*}

\subsection{Overview}
\label{sec:interface-overview}

OpenRoundup's interface, shown in Fig.~\ref{fig:interface}, is designed for laptop and desktop displays and follows standard design patterns for visual analytics systems~\cite{weaver-building-2004}. Four persistent panels are always visible: the \interfaceName{Data Inventory Panel}, the \interfaceName{Schema Panel}, the \interfaceName{Table Rows Panel}, and the \interfaceName{Composite Schema Panel}. The \interfaceName{Column Focus Panel} opens on demand in response to explicit user action. All panels are resizable, enabling users to allocate screen space according to the demands of their current subtask; visualizations and interactive elements are container-responsive. 

\begin{figure}[t]
  \centering
  \includegraphics[width=0.9\columnwidth]{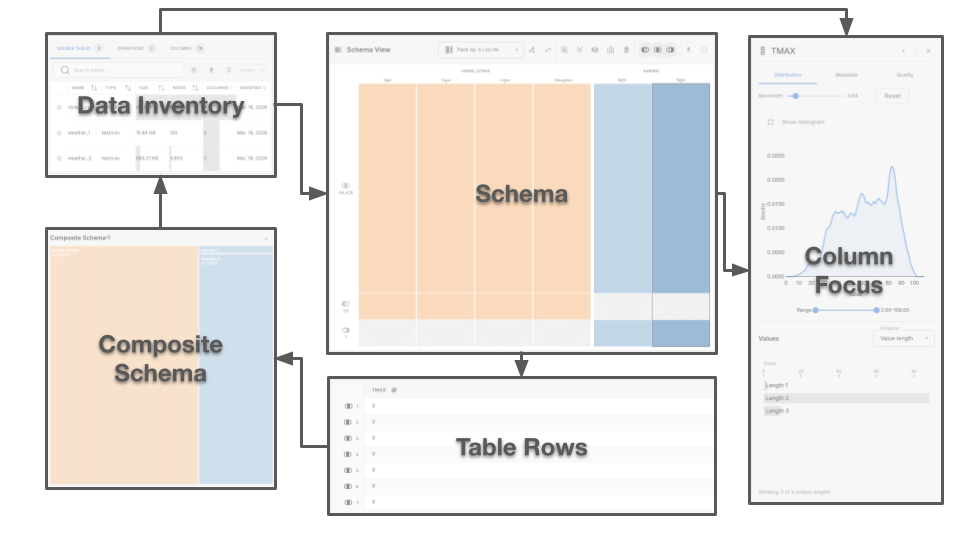}
  \caption{OpenRoundup interaction flow. Users navigate clockwise between four persistent panels: \interfaceName{Data Inventory}, \interfaceName{Schema}, \interfaceName{Table Rows}, and \interfaceName{Composite Schema}. The \interfaceName{Column Focus} on-demand panel appears when columns are focused.}
  \label{fig:interaction-flow}
\end{figure}

The layout is designed for a clockwise interaction flow, as shown in Fig.~\ref{fig:interaction-flow}, which mirrors the four-stage snowball interaction sequence discussed in Sec.~\ref{sec:table-consolidation}. 
Users begin in the \interfaceName{Data Inventory Panel} to \defn{select} tables (S1) and \defn{choose} an operation (S2), move to the \interfaceName{Schema Panel} to \defn{define}  the operation schema (S3), then consult the \interfaceName{Table Rows Panel} and \interfaceName{Composite Schema Panel} to inspect the \defn{materialized} result (S4). The \interfaceName{Column Focus Panel} can be reached as a detour from any data object or from the \interfaceName{Schema Panel} when detailed column-level analytics are needed. This layout implements the Visual Information-Seeking Mantra --- overview first, zoom and filter, then details on demand~\cite{shneiderman-eyes-1996} --- adapted to a pattern appropriate for data integration work through snowballing that we call \defn{schema-first, values-on-demand}, which situates schemas as active throughout every phase of data analysis rather than as a terminal artifact~\cite{berret-iceberg-2024}. The user operates on an incrementally changing schema that is continuously visible, and materializes it to inspect values when desired. 

Throughout this section, we describe aspects of the interface design using Norman's action-oriented vocabulary~\cite{norman-design-2013}: affordances, signifiers, constraints, mappings, and feedback. Many aspects of the interface follow an \defn{object-action interaction model}: the user first selects one or more objects, then invokes an action upon those objects. We chose this model due to its utility for direct manipulation~\cite{shneiderman-designing-2016}, as opposed to the inverse model where the user specifies actions first and then selects objects next. 
Toolbars and menus across the interface implement constraint-based feedback via disabled state, which is both visually signalled and functionally inhibited, to prevent invalid actions at the interface level. The disabled state functions simultaneously as a constraint and as a signifier, communicating to the user which actions are currently available given the state of selected objects.

OpenRoundup interactions are built around three types of objects: Tables, Columns, and Operations. Operations can act on other operations Tables are composed of Columns, and Operations These have a hierarchical relationship: 

\subsection{Data Inventory Panel}
\label{sec:data-inventory}


The multi-faceted \interfaceName{Data Inventory Panel} provides a browsable inventory of all data objects in the workspace via three 
tabs: Source Tables, Columns, and Operations. The tab structure decouples these three central types of objects from the hierarchical linkages of the operation tree, offering direct access to each type without navigating the hierarchy explicitly. The panel expands automatically to the relevant tab to reduce interaction steps and keep views synchronized, for example by switching to the Operations tab when the user creates a new operation, which functions as a feedback signal confirming that the operation was successfully created and directing the user's attention to where it now lives.

\paragraph{Source Tables tab} This tab provides a sortable list view that serves as a data catalog, as shown in \DataInventoryFig. A toolbar provides search and batchable actions above a table of selectable rows containing table metadata. Quantitative metadata fields contain embedded bar visualizations~\cite{goffin-exploring-2014} that encode relative scale at a glance, enabling preattentive value comparisons without leaving the overview. This visual encoding is helpful when comparing quantitative values whose magnitude differences are large enough that the units change, such as file sizes.

\paragraph{Operations tab} The Operations tab (\MegaFigOpsTab) surfaces workflow provenance and mutable operation parameters via a reverse-chronological list rendered as accordion disclosure widgets. Each widget exposes operation metadata with modifiable parameters. Reverse-chronological ordering aligns with the user's working memory and linearizes the operation tree, trading structural expressiveness for direct access and scannability. This linearization is possible because the operation tree is not a branching DAG. Changes to parameters are not applied reactively; the explicit Update button provides a deliberate commit step, functioning as both a constraint that prevents accidental re-computation and a clear mapping between user intent and system action, avoiding triggering expensive re-computations unnecessarily, such as Cartesian products in Pack operations.

\paragraph{Columns tab} As shown in \MegaFigColsTab, this tab presents a flat, sortable table of all columns across all source tables in the workspace, with user-configurable display of columns spanning three groups of properties: identity (column-type adaptive), value statistics (column-type adaptive) and numerical profile (only applicable to numerical type columns). By default, three of the identity columns are shown: Name, Parent, and Type. \SuppFigColDisplay{} shows the full set of options. Sortability affords attribute-level anomaly detection for data quality assessment: sorting by null percentage or unique count maps column position to data quality rank, surfacing problematic columns through spatial ordering rather than requiring the user to scan for them. The three property groups mirror commercial data profiling tool conventions and implement progressive disclosure with respect to column metadata density.

\subsection{Schema Panel}
\label{sec:schema-panel}

The \interfaceName{Schema Panel}, \SchemaFig, is the central interactive region of the interface, displaying the schema of whichever table or operation is currently in focus, so that users can examine intermediate table representations during wrangling. We use a single panel with context-sensitive rendering, as opposed to separate panels per object type, to enforce spatial consistency. The panel comprises a context-sensitive schema toolbar, where actions unavailable for the current object type are disabled, and a schema body displaying the direct descendants of the focused object in the operation tree. Displaying only direct descendants addresses visual scalability as the operation tree grows.

When either a table or operation is in focus, the \interfaceName{Schema Panel} supports contiguous multi-click selection of column partitions within the schema followed by toolbar bulk actions. Both focus contexts implement direct manipulation~\cite{shneiderman-direct-1983} across the schema. Although schema visualizations differ between focused objects, the selection model is consistent across all views: contiguous multi-click selection of column partitions supports bulk actions via the toolbar and targeted on-demand details of corresponding values in the \interfaceName{Table Rows Panel}, reinforcing the connection between schema and values. The toolbar provides context-sensitive actions for the focused object type, for example column-level actions when a table is focused. The consistent selection model across focus contexts reduces the learning burden by establishing a stable mapping between the selection gesture and its effect regardless of what is focused. One action available in all focus contexts is to hide selected columns, which reduces visual complexity without deleting the underlying data.

\paragraph{Column cards} The column card responsive component has a fixed header showing semantic metadata and a body that adapts to available vertical space. It is the atomic unit of the \interfaceName{Schema Panel}. The card boundary and header afford selection via click, and the drag cursor signifies reorderability. When maximally compressed, the card only contains metadata displayed in the header, as shown in \MegaFigTableSchemaCompressed. At an intermediate height, the card body renders top values as labeled pill badges, conveying value identity without magnitude, as shown in \MegaFigTableSchemaCondensed. When more height is available, it switches to proportional horizontal bars encoding both values and relative frequencies, providing a richer mapping between visual extent and value frequency, as shown in \SchemaFig. Cards are color-coded by parent operation using a shared palette, applying pre-attentive color encoding~\cite{ware-information-2004} as a signifier of column-to-table-to-operation provenance across the entire schema. We prioritized displaying value content because it provides the strongest signal for column semantics; metadata such as the column names may convey only limited meaning.

When a table or stack operation is focused, column cards are interactive: clicking a card focuses the corresponding column, and shift-clicking initiates contiguous multi-card selection. Clicking and dragging a card onto another swaps their positions within the table; the drag gesture affords reordering, and the drop target highlight provides feedback confirming where the card will land. Users are not permitted to drag columns between tables; this constraint is enforced by restricting valid drop targets to positions within the same table, preventing a structurally invalid operation at the interaction level.

\paragraph{Table focus} When a source table is focused, the \interfaceName{Schema Panel} renders a single scrollable horizontal row of column cards, as shown in \MegaFig{d-e}.
This layout reflects OpenRoundup's conceptualization of a source table as an ordered set of columns: the column is the primary unit of structure. Column cards are draggable within the row, enabling users to reorder columns before stacking --- where column index determines how tables are aligned in the composite.

\paragraph{Stack focus} When a Stack operation is focused, the \interfaceName{Schema Panel} arranges column cards in a table-partitioned matrix, as shown in \SchemaFig. This view vertically arranges source tables and horizontally arranges column positions corresponding to indices within those tables. Hence, each matrix cell corresponds to an individual child table's column, but the entire vertical vector of columns represents a table-partitioned view of the local composite table's column. When tables have different column counts, the matrix is padded with empty cells to match the widest table, making schema misalignment perceptible through the visual mapping of empty cells to missing columns. Child table columns are automatically concatenated into the output column, without the user needing to reconcile column names or types (see Sec.~\ref{sec:architecture}). 

Index-based alignment is a deliberate design choice grounded by the problem of diachronic data~\cite{kasica-dirty-2023}, data periodically published over time. Semantically equivalent columns tend to recur at stable positions and alongside stable neighbors across releases. Our interaction model honors that expectation: columns are draggable within each table row, enabling the user to correct \defn{schema drift}~\cite{kandel-enterprise-2012}, incremental changes to the table schema across releases. This small multiples design~\cite{tufte-envisioning-1990} shares a coordinate space that maps to the semantics of a union, making schema misalignment across tables perceptible through visual comparison~\cite{gleicher-visual-2011}.

\paragraph{Pack focus} When a Pack operation is focused, as shown in \MegaFigPackSchema, the \interfaceName{Schema Panel} renders a block grid partitioned vertically into three groups of output rows: rows matched between both tables, unmatched rows in the left-hand table, and unmatched rows in the right-hand table. These categories correspond to the three-item toggle button group in the toolbar, establishing a direct mapping between the spatial grouping visible in the schema and the toggle controls that filter those groups. The toggle design enables OpenRoundup to support eight different join types, including anti-joins, as enumerated in \SuppSecJoins. Venn diagram icons serve as signifiers of join type, prioritizing familiarity over formal precision.

As shown in \MegaFigPackSchema, crosshatch texture marks null areas of the output in both partitions of unmatched rows, signifying the absence of values without requiring the user to inspect individual cells. Color coding and a vertical divider between the two tables' column sets apply Gestalt principles to communicate provenance without labeling every cell.

\paragraph{Alerts} OpenRoundup surfaces schematic errors and warnings to provide guidance and feedback to the user while consolidating tables. Alerts are surfaced primarily in the schema toolbar, providing feedback at warning or error severity in direct response to the current operation's state, as shown in \SchemaFig. Stack operation alerts check that child tables have compatible schemas for stacking: incongruent column counts raise an error, and mismatched types at the same index raise a warning. Pack operation alerts check that join keys, join type, and a matching predicate have all been specified, functioning as constraints that surface incomplete configuration before the user commits to materializing a result.

\begin{figure*}[t]
  \centering
  \includegraphics[width=1\textwidth]{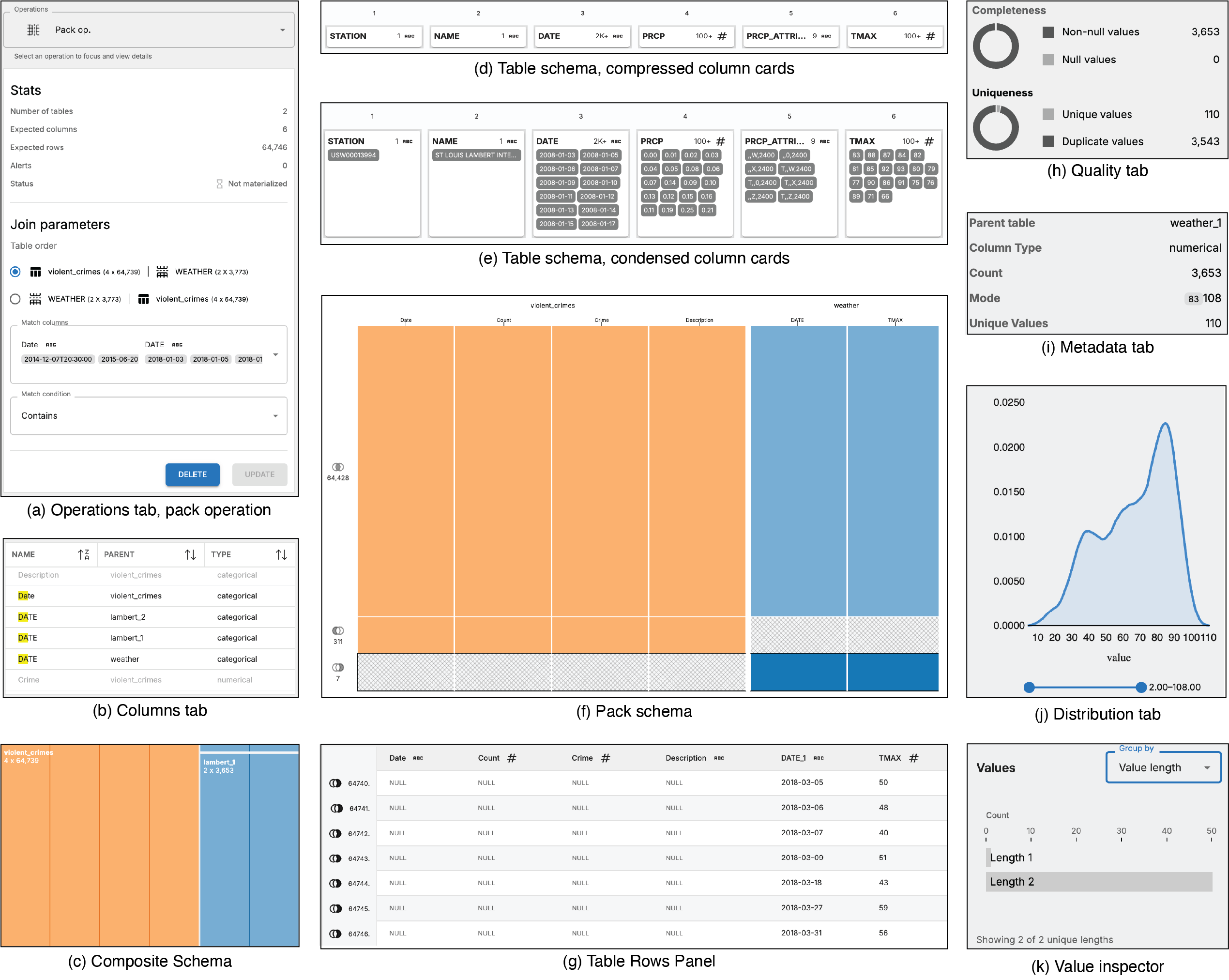}
  \caption{Transient views: alternative tabs and states. The \interfaceName{Data Inventory Panel}: (a) the Operations tab showing a pack operation; (b) the Columns tab. (c) The \interfaceName{Composite Schema Panel} after a pack operation. Responsive column cards in the \interfaceName{Schema Panel} view: (d) Compressed (e) Condensed. (f) The \interfaceName{Schema Panel} after a pack operation, with bottom right partition selected. (g) The \interfaceName{Table Rows Panel} showing that selected data. The \interfaceName{Column Focus Panel} in Detail mode when the single \texttt{TMAX} column is in focus has three possible tabs in the top half: (h) quality, (i) metadata, (j) distribution; (k) the bottom half shows Values. 
    }
  \label{fig:megafig}
\end{figure*}

\subsection{Table Rows Panel}
\label{sec:table-rows}

The \interfaceName{Table Rows Panel}, shown in \TableRowsFig and \MegaFigTableRows, is a selection-driven tabular inspection panel for understanding cross-column value relationships. Tabular artifacts are ``inextricably enmeshed with sensemaking''~\cite{bartram-untidy-2022}, and database users consistently prefer row information and tabular interfaces~\cite{kahng-interactive-2016}.  The panel displays rows of the focused table or materialized operation, restricted to whichever columns are currently selected in the \interfaceName{Schema Panel}. The dependency on an explicit column selection instantiates a constraint that enforces the schema-first, values-on-demand principle: no columns selected means no rows displayed, preventing the cognitive burden of scanning wide tables before the user has identified which columns are relevant. Columns are sortable, and additional rows are loaded incrementally via infinite scroll. Columns are annotated in margins with icons specific to the operation type.

When no rows are displayed, there are four distinct empty states, each of which provides targeted feedback to orient the user toward the correct next action depending on the underlying reasons: no columns are selected, there are errors in the schema specification, the operation is unmaterialized, or the materializations operation is out of sync with the schema specification. This specificity transforms an absence of data into an actionable signifier of the required next step, which would not be the case if there were only a generic placeholder. 

\subsection{Composite Schema Panel}
\label{sec:composite-schema}

The \interfaceName{Composite Schema Panel}, shown in \CompositeSchemaFig~and \MegaFigCTS, is a recursive treemap visualization~\cite{ahlberg-dynamic-1992} of the entire operation tree, updated in real time as operations are created or modified, providing continuous feedback on the evolving structure of the workflow. Each table block is subdivided into equally-spaced column ticks and sized vertically in proportion to its row count relative to sibling tables within the same operation, establishing a mapping between visual area and row count that gives immediate perceptual weight to larger tables. Table names and dimensions are displayed when a block is large enough to accommodate them. Stack operations arrange these table blocks vertically, and Pack operations arrange them horizontally. This spatial mapping between layout axis and operation semantics~\cite{ziemkiewicz-shaping-2008} communicates operation type without requiring labels. It was more effective than all the alternative hierarchical visualization approaches we considered, including node-link diagrams and icicle plots, because the proportional breakdown directly reflects the underlying table structure.  

Proportional row-count sizing signifies which source tables dominate the output, a critical signal when diagnosing unexpected output sizes after stacking or packing. Column ticks are minimal signifiers, conveying only presence, count, relative position, and interaction state, supporting structural comparisons across tables at the overview level. Tracking properties of tabular data across many operations is cognitively demanding~\cite{shrestha-unravel-2021}, so the \interfaceName{Composite Schema Panel} is designed to orient the analyst throughout the assembly process.

\subsection{Column Focus Panel}
\label{sec:column-focus}

The \interfaceName{Column Focus Panel} is an on-demand panel for detailed column-level profiling. Focusing one to five columns from either the \interfaceName{Schema Panel} or the Columns tab of the \interfaceName{Data Inventory Panel} opens the panel as a drawer; the blurring of non-focused columns in other views functions as a signifier of the focused state, communicating which columns are under active inspection and constraining attention to the selected set. The panel operates in either detail or comparison mode depending on the number of focused columns.

\paragraph{Detail view (one column)} A tabbed overview provides three perspectives whose tab labels signify the analytical question each addresses: a Distribution tab (numeric columns only) showing a KDE plot with an adjustable bandwidth slider, bi-directional domain filter, and an optional histogram overlay (\MegaFigDistributionTab); a Metadata tab providing value lookups for name, type, parent table, count, mode, unique count, and numeric summary statistics (\MegaFigMetadataTab); and a Quality tab (\MegaFigDQTab) showing completeness and uniqueness as donut charts~\cite{skau-arcs-2016}. These two metrics are chosen for their general applicability across column types and their direct relevance to data integration tasks. 

We chose a KDE plot over a histogram for distributional overview because it surfaces the continuous shape of the distribution without perceptual artifacts from arbitrary bin boundaries~\cite{correll-looks-2019}. The bandwidth slider affords adjustment and makes the smoothing parameter an explicit, user-controllable analytical decision rather than a hidden default; Silverman's rule of thumb provides a principled starting point~\cite{silverman-density-1986} that maps to a sensible initial state.

Below the tab panel, as shown in \MegaFigValueCounts, a scrollable value count distribution displays the frequency of each unique value as a horizontal bar chart, groupable by value or by value length, which serves as a data quality diagnostic for detecting anomalous lengths.

\paragraph{Column comparison mode (two to five columns)} As shown in \ColumnComparisonFig, a summary table lists each focused column with identification metadata including parent table and column index, disambiguating provenance, which is a prerequisite for comparing columns with the same name. A metadata section renders bar charts comparing aggregate statistics, with a box-and-whisker plot when all focused columns are numeric. We chose it for this context due to its compactness and the fact that the required statistics are pre-computed when columns are instantiated in the system (see Sec.~\ref{sec:architecture}), so the plot can be rendered without additional computation.

The majority of the panel is occupied by a value UpSet visualization~\cite{lex-upset-2014}: a scrollable matrix where each row is a unique value present in at least one focused column, filled circles signify presence in that column, and connecting lines signify shared values across multiple columns. Values can be filtered by text input and sorted by degree or alphabetically. We chose an UpSet plot over a Venn diagram for scalability and value identity preservation: it affords asking not just how much overlap exists, but which exact values are shared and which are unique to one source. This value-level comparison supports the join key qualification task at the heart of table consolidation.

\subsection{View Coordination}
\label{sec:view-coordination}

Color (hue) encodes direct parent operation across all views, cycling when the number of operations exceeds the palette. Highly saturated red is reserved for error states due to its visual salience, functioning as a signifier of severity that is consistent with convention and pre-attentive color perception~\cite{ware-information-2004}. Extensive linked highlighting and selection are maintained consistently across all panels~\cite{qu-keeping-2018} via shared UI state following the multiple coordinated views paradigm~\cite{north-snap-together-2000}, as illustrated in \SuppFigMCVDetails. This cross-panel feedback ensures that every selection in one view produces an immediate, visible response in all others, reinforcing the mapping between data objects and their representations: selecting a column in the \interfaceName{Schema Panel} displays it in the \interfaceName{Table Rows Panel} (schema-first, values-on-demand), and selecting or hovering columns in the \interfaceName{Schema Panel} updates styling in the \interfaceName{Composite Schema Panel}, helping analysts trace individual columns through the full workflow structure.

\section{Replication Study and Usage Example}
\label{sec:usage}

To demonstrate OpenRoundup's expressiveness, the first author replicated 17 table consolidation workflows originally published as programming notebooks by working journalists. The source workflows were drawn from 50 publicly available GitHub repositories written in a scripting language and analyzed extensively in previous work~\cite{kasica-table-2021}. From that dataset, these 17 met the inclusion criteria: consolidates tables, uses tabular data, and the raw source data is obtainable. 

Table~\ref{tab:datasets} summarizes these 17 workflows, which cover a wide range of story topics including refugee resettlement, violent crime and heat, voter registration, campaign finance, IRS audit rates, drug payments to doctors, and partisan baby name analysis. The replicated workflows span 2 to 51 source tables and use Stack alone, Pack alone, or both operations in combination. Below we present one workflow in detail as a usage example. See \SuppSecWorkflows{} for the other 16 replicated workflows at varying levels of detail, including source links, descriptions, and screenshots. This workflow and two others can also be seen in the supplemental videos. 

We use the term \defn{usage example} rather than \textit{usage scenario} to signal that the walkthrough below follows a concrete, published workflow, not a hypothetical task imagined by visualization researchers. The walkthrough illustrates how OpenRoundup supports the snowball interaction sequence across all four phases of table assembly. 

\begin{table}[t]
\centering
\caption{The 17 journalism data workflows replicated using OpenRoundup to demonstrate its expressiveness, drawn from publicly available GitHub repositories analyzed previously~\cite{kasica-table-2021}.}

\label{tab:datasets}
\begin{tabular}{p{0.35\linewidth} p{0.06\linewidth} p{0.2\linewidth} p{0.1\linewidth}}
\toprule
\textbf{Topic} & \textbf{Tables} & \textbf{Publication} & \textbf{Category} \\
\midrule
\midrule
\textbf{Crime \& heat correlation} & 3 & \textit{STLPR} & Hybrid \\
Campaign donations & 17 & \textit{BuzzFeed News} & Hybrid \\
IRS audit patterns & 5 & \textit{ProPublica} & Hybrid \\
Elder care abuse & 8 & \textit{The Oregonian} & Hybrid \\
Public school performance & 4 & \textit{Baltimore Sun} & Hybrid \\
Name partisanship & 51 & \textit{Time} & Hybrid \\
\midrule
Energy access inequality & 5 & \textit{WUFT} & Pack \\
School vouchers & 2 & \textit{NPR} & Pack \\
Infrastructure job growth & 4 & \textit{FiveThirtyEight} & Pack \\
Refugee resettlement & 2 & \textit{BuzzFeed News} & Pack \\
Election results & 10 & \textit{Times} & Pack \\
\midrule
Agricultural worker visas & 9 & \textit{LA Times} & Stack \\
Rideshare analysis & 4 & \textit{Quartz} & Stack \\
Big Pharma payments & 12 & \textit{CORRECTIV} & Stack \\
Demolition permits & 2 & \textit{The Statesman} & Stack \\
Voter registration & 11 & \textit{Baltimore Sun} & Stack \\
Workplace discrimination & 8 & \textit{CPI} & Stack \\
\bottomrule
\end{tabular}
\end{table}

\subsection{Crime \& Heat: A Detailed Walkthrough}
\label{sec:crime-heat}

The workflow investigates whether violent crime rises during hot weather in St.\ Louis, Missouri, an instance of exploratory multi-table data wrangling in which the journalist begins with a hunch that may or may not yield a publishable story. Three source files are used: \texttt{weather\_1.csv} (3,653 rows $\times$ 45 columns; daily temperature records 2008--2017), \texttt{weather\_2.csv} (120 rows $\times$ 21 columns; daily records for 2018), and \texttt{violent\_crimes.csv} (64,739 rows $\times$ 4 columns; St.\ Louis Metropolitan Police Department incident file, 1967--2018). The goal is a single table linking the count of violent crime incidents to maximum temperature for each day over the last decade. The story was published in 2018.

\paragraph{Upload and initial exploration} The user uploads all three files via the browser's native file upload interface. The Source Tables tab of the \interfaceName{Data Inventory Panel} populates with name, type, size, row count, column count, and modification date for each table, with embedded bar marks, as shown in \DataInventoryFig. The user observes that \texttt{violent\_crimes} has many rows but few columns, that \texttt{weather\_1} is much larger than \texttt{weather\_2}, and that the two weather tables have different column counts.

Focusing \texttt{weather\_1} shows it in the \interfaceName{Schema Panel}, and the user identifies a \texttt{DATE} column as a primary key because all values appear exactly once. On the other hand, the first two columns, \texttt{STATION} and \texttt{NAME}, have only one unique value each, suggesting they are constant across all rows and hence not useful for analysis, as visible in \MegaFigTableSchemaCondensed. These columns along with other irrelevant columns could be removed from the schema without affecting the integrity of the data.

The user focuses on the \texttt{TMAX} column and sees it in the \interfaceName{Column Focus Panel}. From the Distribution tab, the user can see that the distribution does not have any anomalous spikes or outliers, as shown in \MegaFigDistributionTab, and the user can verify the range of values via the filter slider below the x-axis that shows $[2, 108]$. By selecting the Metadata tab, the user can confirm other statistics, including the total number of values, the most frequently occuring, and the number of unique values, as shown in \MegaFigMetadataTab. These values are in an expected range for Fahrenheit for the local climate, which is information not present in the metadata. As shown in \MegaFigDQTab, the user confirms that the column is \textit{complete}, there are no null values, in the data Quality tab.


\paragraph{Stacking the weather tables} The user selects both weather tables in the \interfaceName{Data Inventory Panel} and creates a Stack operation from the Actions dropdown. In response, OpenRoundup creates a new operation node at the operation tree root, links selected weather tables to the newly created operation, and updates the \interfaceName{Schema Panel} to show a table-partitioned matrix, as shown in \InterfaceFigRef{b-d}. 
The toolbar raises an alert because the two weather tables have different column counts. To remove all columns except for the two columns relevant to their inquiry, \texttt{DATE} and \texttt{TMAX}, the user clicks and drags these columns in their respective tables to the leftmost two indices, which are shared across both tables. During this process the system raises a warning-level alert because columns at the shared index have heterogeneous types: one column is categorical, the other is quantitative. The user resolves this mismatch by dragging the second \texttt{TMAX} column to the same index as the first, aligning the previous index with columns of homogeneous types. To delete the other columns, the user shift-selects all columns to the right and clicks the delete button in the \interfaceName{Schema Panel} toolbar. After a confirmation dialog prompt, the results are visible. With the schema specified, the user renames the operation \texttt{weather} for clarity.

As one final diagnostic check, the user selects both \texttt{DATE} columns and opens the \interfaceName{Column Focus Panel} in comparison mode: the UpSet visualization shows they form disjoint sets, as expected for non-overlapping date ranges. Selecting both \texttt{TMAX} columns (\ColumnComparisonFig), the box-and-whisker comparison confirms distributional differences consistent with year-to-year variation but semantic equivalence.

Satisfied, the user materializes the stack. The \interfaceName{Table Rows Panel} confirms one row per day across the full multi-year span, with \texttt{DATE} formatted as \texttt{YYYY-MM-DD} strings.

The \interfaceName{Composite Schema Panel} (\CompositeSchemaFig) shows the full operation tree with the \texttt{weather} stack operation as the root and the two weather tables as its children. The vertical sizing of the table blocks within the stack reflects their relative row counts: \texttt{weather\_1} is 30X the size of \texttt{weather\_2}.

\paragraph{Packing the crime table} With the \texttt{weather} operation in focus, the user selects the \texttt{violent\_crimes} table from the \interfaceName{Data Inventory Panel} (\DataInventoryFig), and selects the pack table options from the Actions dropdown. OpenRoundup creates a new operation node and focuses it. The \interfaceName{Schema Panel} displays the pack schema partitioned by source table and row-match category. An error-level alert fires because join parameters have not yet been specified. 

In the Operations tab, the user opens the "Match columns" combobox, as shown in \MegaFigOpsTab. The list presents ranked column pair candidates, surfacing \texttt{DATE}/\texttt{Date} as the top choice. The user selects this pair, commits the changes by pressing Update, and observes the result in the \interfaceName{Schema Panel}: zero matching rows, with all \texttt{weather} rows in the left-unmatched group and all \texttt{violent\_crimes} rows in the right-unmatched group. Inspecting example values in the combobox, the user discovers that \texttt{DATE} uses \texttt{YYYY-MM-DD} formatting while \texttt{Date} includes time (\texttt{YYYY-MM-DDThh:mm:ss}). The user switches the match predicate to \texttt{CONTAINS}, testing whether the crime datetime string contains the weather date string as a substring, and commits the changes. The \interfaceName{Schema Panel} updates: 64,000$+$ matching rows, 311 left-unmatched weather days, 7 right-unmatched crime rows, as shown in \MegaFigPackSchema. The user materializes the operation, and sees that the \interfaceName{Composite Schema Panel} updates, adding four orange columns representing \texttt{violent\_crimes} to the structural overview (\MegaFigCTS).
  
\paragraph{Inspection and export} Examining the unmatched rows in the \interfaceName{Table Rows Panel}, the user determines that the unmatched weather days fall at the beginning of March 2018 and the unmatched crime rows fall outside the 2008--2018 date range, as shown in \MegaFig{f-g}. The user toggles the match-type buttons in the \interfaceName{Schema} toolbar to deselect left-only and right-only groups, effectively performing an inner join. The system raises an out-of-sync alert: the specification of the schema has changed since the operation was last materialized. The user re-materializes the operation, which clears the alert.

The final export produces a CSV integrating crime incidents with daily maximum temperatures, ready for downstream analysis such as grouping rows by date, summing crime counts, and computing maximum temperature per day. These tasks would be performable in Excel, Python, R, or a declarative visualization grammar such as Vega-Lite~\cite{satyanarayan-vega-lite-2017}.

\section{Architecture}
\label{sec:architecture}

\begin{figure}[h]
  \centering
  \includegraphics[width=.7\columnwidth]{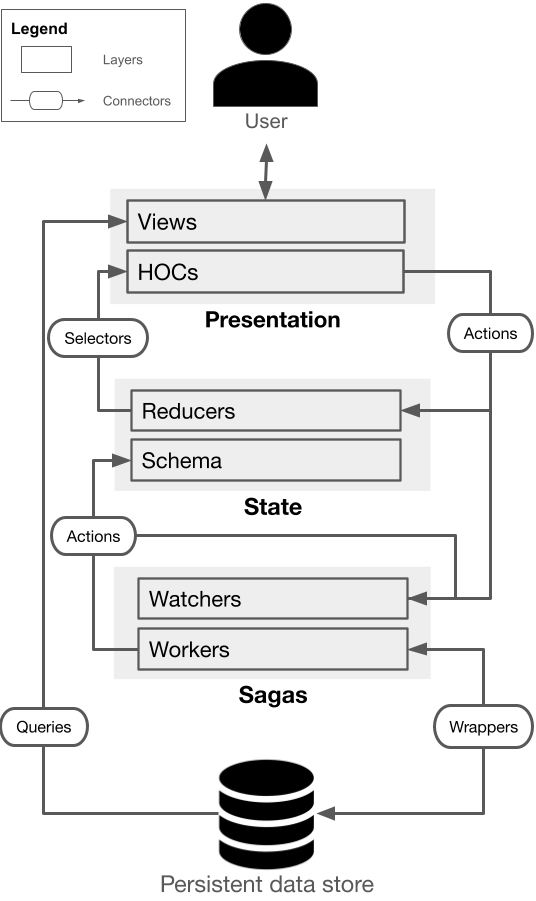}
  \caption{OpenRoundup is built on a three-layer architecture that mediates interactions between the user and persistent data stores, with four connectors binding the layers together.}
  \label{fig:architecture}
\end{figure}


OpenRoundup is built on a multi-layer architecture that mediates interactions between the user and persistent data stores, with domain abstractions centered on columns, tables, and operations. The architecture supports the complex, multi-step workflows characteristic of table consolidation tasks while maintaining a clear separation of concerns and enabling future extensibility. It comprises three layers maintaining dual representations of the application's data: Presentation, UI components and visualizations; State, structured domain objects; and Sagas, stateless transformation and coordination. Four connectors bind the layers: Actions, Selectors, Wrappers, and Queries.

\paragraph{Persistent data stores} While not an explicit layer of OpenRoundup's architecture, persistent data stores are a dependency. These external systems store and transform raw data on behalf of the system. This design choice decouples OpenRoundup's core logic from specific backend implementations, enabling future support for additional backends and federated data sources (see Sec.~\ref{sec:discussion}) without requiring changes to the Presentation or State layers. OpenRoundup interacts with them exclusively via Wrappers from the Sagas layer. The current implementation uses DuckDB-WASM~\cite{kohn-duckdb-wasm-2022}, an in-browser columnar OLAP engine compiled to WebAssembly, chosen for its virtual filesystem support, worker-thread isolation, and ability to execute standard SQL client-side. 

\subsection{Layers}
\label{sec:layers}

\begin{figure}[h]
  \centering
  \includegraphics[width=\columnwidth]{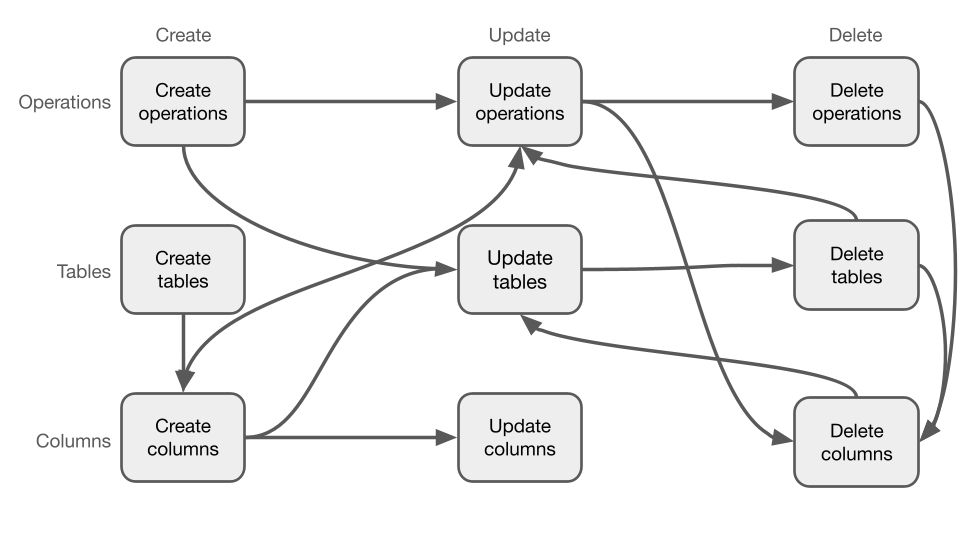}
  \caption{This $3 \times 3$ matrix defines the nine sagas used to transform data in OpenRoundup.}
  \label{fig:saga-matrix}
\end{figure}

\paragraph{Saga} The \defn{Saga layer}, the closest architectural layer to the raw tables in persistent data stores, is a stateless, bidirectional translation layer that constitutes the sole pathway between State and a persistent data store. This layer is composed of nine sagas, each corresponding to one of nine operations defined by a $3 \times 3$ matrix of three operation types (create, update, and delete) on three data objects (columns, tables, operations), as illustrated in Fig.~\ref{fig:saga-matrix}. 

Garcia-Molina \& Salem~\cite{garcia-molina-sagas-1987} define a \defn{saga} as a long-lived database transaction decomposed into a sequence of sub-transactions $T_1, T_2, \ldots, T_n$, each paired with a compensating transaction $C_i$ that semantically undoes $T_i$'s effects. The system guarantees either full completion or full compensation, never a silent partial execution. OpenRoundup's Saga layer borrows this decomposition metaphor, breaking complex, multi-step persistence operations into discrete, independently observable steps. However, it does not implement the compensation mechanism. The client-only architecture and single persistent data store make crash-recovery guarantees unnecessary in practice. 

Each saga in the Saga layer is composed of two sub-components: a Watcher and a Worker. \defn{Watchers} listen for Actions and invoke Workers. \defn{Workers} execute atomic operations and dispatch result Actions back to State. This separation of concerns allows Watchers to focus on complex conditional logic and orchestration, while Workers remain simple and focused on executing discrete persistence operations. The separation of these two responsibilities is load-bearing: Watchers listen for success signals rather than for the domain slice actions, which keeps cascade subscriptions decoupled from the slice's internal update events. Workers have no knowledge of which Watchers, if any, will respond to their success signal. Cascade effects are emergent from independent Watcher subscriptions rather than orchestrated by the Worker.

Watchers are logic-heavy by design. All conditional reasoning is concentrated here. A watcher listens for system dispatched events, specifically \textit{actions}, and queries the global data state, represented in State, to establish context before invoking exactly one Worker with appropriate parameters. Handling specific events and data state context together is what allows OpenRoundup to perform expressive table transformations with a small set of sagas. For example, inserting a column and creating a column use the same system event (the column creation \textit{action}), but in the former case a table will already have columns and in the latter it will not, a distinction maintained in \textit{State}.

Watchers' cascading behavior is emergent from the interconnected network of Watcher-Action relationships, similar to a publish-subscribe pattern. Multiple independent Watchers can listen for the same Action, while cascading side effects arise without centralized orchestration. Some cascades are also recursive. For example, inserting a column into a stack operation requires updating child tables or operation columns. Watchers also handle these kinds of recursive updates before passing the payload to the appropriate worker.

Workers are simple and focused. A Worker receives a payload comprising a batch of items, then processes them in sequence, accumulating results and halting on failure (but not recovering). Dispatch is atomic at the batch level: either all items in a batch succeed and their results are committed to the global data state, or no results are committed. Each Worker dispatches exactly two system events, \textit{actions}, on success: a domain slice action that updates the global data state, and a named success action prefixed by the saga, for example \texttt{createTablesSuccess}, that may trigger downstream cascade behavior. 

\paragraph{State} The \defn{State layer} maintains a global, semantic representation of application status via five bounded domains, called \defn{slices}. We divide these into two groups: \defn{data slices}, which represent the global data space, and \defn{UI slices}, which represent the global interaction state.

The \defn{global data space} is composed of three slices: columns, tables, and operations. Each of these slices itself is structured using a normalized, bidirectionally linked structure. We explored other ways of representing global data space, including node-link structures and matrices of columns, but settled on normalized, bidirectionally linked slices for their simplicity, flexibility, and efficiency.

Each object is uniquely identifiable via an auto-incrementing object-type prefixed ID, for example the first table in the system would have an ID of \texttt{t1}.  This convention not only produces universally unique identifiers, similar to a primary key, but this prefix supports O(1) object type discrimination through the OpenRoundup system. With these IDs tables link between columns and operations, columns link to their parent tables or operations, and operations link to their child tables and columns. Bidirectional links between all related objects support O(1) lookups by ID and O(n) ordered iteration across all views.

The \defn{Columns slice} stores pre-computed statistical metadata: name, column type, null counts, unique count, top values, numeric summary statistics (if applicable for the column type), and linkages. We store type information at this level conceptually and structurally. We enforce that a column has a single homogeneous type, in contrast to other systems such as OpenRefine. We also do not manage column type information in the persistent data store. Type information drives user-facing diagnostics and interaction affordances, for example range sliders for numeric columns and text input for categorical ones. 

The \defn{Tables slice} stores provenance and structural metadata for each uploaded source table. Each Table object records display identity (\texttt{name}, mutable) separately from file identity (\texttt{fileName}, \texttt{extension}, \texttt{mimeType}, \texttt{dateLastModified}, \texttt{size}), preserving the original metadata even after a user renames a table. A table's \texttt{parentId} records its parent operation in the operation tree, enabling post-order traversal. A \texttt{databaseName} field stores the corresponding persistent data store identifier, functioning as a foreign key. Row count is cached as a derived attribute after ingestion, avoiding repeated asynchronous queries for display purposes at the cost of potential staleness, an acceptable trade-off since source tables are immutable after upload.

The \texttt{columnIds} field is an ordered array of column ID references rather than embedded column objects, maintaining column ordering as application state while keeping the slice normalized. Reordering column IDs in this array updates the column order for that table across all views without affecting the underlying column object. The persistent data store is not updated with column order changes, since column order is a presentation concern.

The \defn{Operations slice} stores transformation nodes in the operation tree. Each operation shares a set of common attributes, including name, operation type, and status flags to determine if the operation is materialized and if the schema is in sync with the last materialization. An operation can be materialized but out of sync when the user changes a join key, column ordering, or child table/operation ordering after a prior materialization. These two flags separate the question of \textit{existence} in persistent data stores from the question of \textit{currency}, enabling the UI to distinguish between ``never materialized'', ``materialized and current'', and ``materialized but stale'' without querying persistent data stores.

The Operations linkage array to child objects is heterogeneous: it may contain table IDs, operation IDs, or both, since operations can be composed into a tree. The slice also tracks a single \texttt{rootOperationId}: every newly added operation unconditionally becomes the new root, which enforces the snowball assembly model in State.

Pack and Stack operations have different object schemas within the same slice. When \texttt{operationType} is \texttt{PACK}, the factory conditionally appends join parameters (\texttt{joinKey1}, \texttt{joinKey2}, \texttt{joinType}, \texttt{joinPredicate}) and a \texttt{matchStats} object recording the matched / left-unmatched / right-unmatched row counts. The join predicate is not limited to equality: \textsc{Contains}, \textsc{StartsWith}, and \textsc{EndsWith} support substring and prefix matching, which is how the crime and heat workflow joins date strings of differing formats without a transformation step. \texttt{matchStats} is a derived attribute populated from persistent data stores after each materialization and cached in State for immediate display without a subsequent query.

The global interaction state is composed of two slices. The \defn{UI slice} holds global interaction state decoupled from domain data: multi-ID buffers for hovered and selected objects, and single-ID indices for focused objects, which power multiple coordinated views. This separation centralizes transient interaction semantics while preserving pure reducer updates in domain slices. The \defn{Alerts slice} stores application-level validation and notification state using the same two-index structure, with each alert recording identity, severity (error or warning), passing status, silence state, and timestamp. Separating alert semantics from both domain data and transient UI state supports composable validation pipelines and a stable contract for user-facing diagnostics.

\paragraph{Presentation} The \defn{Presentation layer} is the user-facing layer. It comprises two sub-layers: views and higher-order components. 

\defn{Views} form a pure presentation layer: UI components concerned only with layout, interaction events, and visual encoding. These views are stateless with regards to the global data model stored in the State layer. They do, however, maintain a local interaction state for transient UI semantics, such as open/closed state of dropdowns and drawers. This sub-layer has no direct State access. All data is passed to these components via the second sub-layer, higher-order components. 

\defn{Higher-Order Components (HOCs)} form a data injection layer: each HOC accepts a single object ID and injects the full domain object, derived metadata, interaction state, and action dispatchers into a plain component. HOCs are composable and functorial. Explicit separation of data injection from presentation improves testability and maintainability.

Cross-cutting patterns include windowed scrolling for large row sets, and reusable primitives (EditableText, DropZone, DescriptionList, themed components).

\subsection{Connectors}
\label{sec:connectors}

\paragraph{Wrappers} Wrappers are the exclusive interface between Sagas and a persistent data store. Each \defn{Wrapper} is a JavaScript module conforming to a stable contract supporting five operation classes: Ingest (load external data), Inspect (retrieve schema and summary statistics without returning row data), Retrieve (paginated row-level access), Transform (create derived artifacts such as joins and unions), and Dispose (remove artifacts). The Wrapper contract enables pluggable backends: Saga Workers call different Wrappers using a common signature based on table metadata, with the correct Wrapper determined by the Watcher rather than the Worker, keeping Workers decoupled from backend selection logic.

\paragraph{Actions} \defn{Actions} are discrete, serializable events representing the complete set of permissible state transitions, enforcing strict unidirectional data flow. Two classes exist: \defn{terminal actions} (Presentation to State) whose effects are fully realized upon reaching a reducer, and \defn{initiating actions} (Presentation to persistent data stores via Sagas) that serve as triggers for long-running workflows.

\paragraph{Selectors} \defn{Selectors} are memoized, composable derived-state accessor functions forming the boundary between the Presentation and State layers. They are rank-polymorphic (accepting singleton values, arrays, and matrices) and null-safe. Memoization maintains referential stability across multiple coordinated views, preventing unnecessary re-renders in components that share state.

\paragraph{Queries} Occasionally, OpenRoundup needs to obtain data that is not stored in the global state and must query the persistent data stores, such as when retrieving column values for the column details view, or retrieving row data for the table rows view. \defn{Queries} are read-only, ad-hoc, one-off requests.

\subsection{Implementation}
\label{sec:implementation}

OpenRoundup is a client-only web application implemented in JavaScript with no server-side runtime. The client-only architecture provides strong data privacy guarantees by ensuring that source files never leave the user's device, important for journalists handling sensitive data. The application is bundled and served via Vite using the SWC compiler for fast transpilation, deploying as a static asset.

\paragraph{Persistent data} File ingestion is handled client-side via the HTML5 File API. All column values are stored in DuckDB-WASM as \texttt{VARCHAR} to preserve source fidelity and avoid implicit type coercion errors during ingestion. Column type semantics are inferred and maintained in the State layer rather than enforced in the persistent data store.

\paragraph{Sagas} The Saga layer is realized through the Redux-Saga library~\footnote{\url{https://redux-saga.js.org/}}, which uses ES6 generators to sequence asynchronous effects declaratively, enabling Watcher and Worker logic to be expressed as readable, synchronous-style code despite coordinating multiple asynchronous persistence calls.

\paragraph{Presentation} UI components use MUI~\footnote{\url{http://mui.com/}} components, and D3~\footnote{\url{https://d3js.org}} utilities handle scales and client-side formatting, but all visualizations are implemented as React components rendering HTML and CSS, except for the KDE visualization that is rendered SVG. Drag-and-drop is implemented via React \textit{DnD}~\footnote{\url{https://react-dnd.github.io/react-dnd}} with a custom drag layer. Panel resizing uses \texttt{react-resizable-panels}~\footnote{\url{https://react-resizable-panels.vercel.app}}. Typography uses Inter~\footnote{\url{https://fonts.google.com/specimen/Inter}}, a variable font distributed via Google Fonts, enabling responsive weight adjustments without font file overhead. Queries are implemented as React's native hooks, enabling components to subscribe to on-demand persistent data reads without Saga mediation.

\paragraph{Testing} Testing uses Cypress~\footnote{\url{https://www.cypress.io}} for verifying component state against application state and Vitest~\footnote{\url{https://vitest.dev}} for architecture testing, particularly integration testing of Saga workers. 

\section{Deployment Study}
\label{sec:deployment}

We designed OpenRoundup to replicate the same table consolidation tasks as those performed by data journalists with programming expertise, as demonstrated in Sec.~\ref{sec:usage}. To evaluate the system with its target users, we deployed it in the wild with professional data journalists, with the primary goal of evaluating the utility of OpenRoundup and the secondary goal of evaluating its usability. 

\subsection{Study Design}
\label{sec:study-design}

We conducted a multi-session, longitudinal deployment study with four professional data journalists. In the first one-hour session we obtained informed consent and background information, demonstrated OpenRoundup, and collected first impressions. Participants then used OpenRoundup for a two to five week tryout period, either working independently with their own data and tasks or with materials hosted on OpenRoundup's documentation site. The site is structured according to the Diátaxis framework~\cite{procida-diataxis-2023} for technical documentation, organizing content into tutorials, how-to guides, reference, and explanation. Following the tryout period, we conducted a second one-hour session with a semi-structured interview covering overall impressions, utility, usability, and incident recall.

We pilot tested our interview scripts with two users outside of journalism. We had six participants for session 1, providing valuable insights, but only four completed the tryout period and session 2. We conducted all sessions remotely via Zoom with no compensation. We recorded audio and video, transcribed the audio, and de-identified these transcripts.

All participants except one also participated in our prior interview study~\cite{kasica-table-2021}. We recruited the additional participant from online professionial journalism organizations (mailing lists and Slack channels). We intentionally sought out participants with varying programming backgrounds, using archetypes of data workers defined by Kandel et al.~\cite{kandel-enterprise-2012} interpreted along a spectrum of literacy with computational tools and techniques. We primarily recruit users who fall between the application user and scripter archetypes, but also include those who fall between the scripter and hacker archetypes, to investigate the utility of OpenRoundup for users with more programming experience. Three participants fell between application user and scripters, with P2 and P4 gravitating more towards the application user end of the spectrum and P1 gravitating more towards the scripter end. P1 fell between the scripter and hacker. See \SuppSecDeployment{} for study materials and participant demographic details.

\subsection{Results: Utility}
\label{sec:results-target}

Across all participants, interview data confirmed that multi-table consolidation is a genuine, recurring challenge in journalistic data work, providing empirical grounding for the problem motivating OpenRoundup's design. P1 characterized the task as a near-daily occurrence: ``every day, you know, like, we're doing that all the time.'' P4 mapped the tool onto an active reporting workflow involving election results, describing the consolidation goal in terms that closely echo our Stack model: ``I'm gonna bash all that together, without doing a lot of programming.'' The strongest corroborating evidence came from independent use during the tryout period, in which participants worked with their own data and tasks without researcher involvement: P2 applied the tool to join school shooting and temperature datasets, and P3 joined tables of churches and church pastors to investigate whether retired pastors were returning to active status. These self-directed tasks demonstrate the system's capability to support ecologically valid consolidation work. Both participants also followed the published tutorials. 

The conceptual framing resonates with target users. P3 described the tool as if to a hypothetical colleague as ``a drag-and-drop way to combine data where you can see what you're doing \ldots\ takes the mystery out of it,'' explicitly contrasting it with code: ``I hope this code works and click run \ldots\ \texttt{VLOOKUP} is just hell.''

OpenRoundup addresses a targeted 
slice of the data preparation space, and our participants concurred that the problem is real. P2 confirmed it is more usable than Excel for multi-column merges---``OpenRoundup would probably be faster, especially once I use it a few more times''---while noting that \texttt{VLOOKUP} remains adequate when only a single column is needed and the source is already an \texttt{XLSX} file. P3 articulated the target audience clearly: ``people who are frustrated by the limitations or annoyances of spreadsheets, of Excel, who want to work with relational data \ldots\ but who do not yet, or maybe never will have, the programmatic skills.'' OpenRoundup is positioned for users who understand what a join is conceptually but lack the coding skills to execute one programmatically; roughly, the intermediate Excel user who has encountered SQL.

An unexpected but strong secondary use case emerged: teaching. P3, who teaches data journalism at the university level, noted: ``every time I teach joins, I literally have to draw on the whiteboard \ldots\ people who are new to this don't really understand what that means until they actually see it,'' and saw the visual structure of the \interfaceName{Schema Panel} in Pack-focus mode as directly addressing this knowledge gap. This observation points to a secondary audience of data journalism educators, who were not anticipated in the original design.

However, participants who fell closer to the scripter and hacker archetypes, recruited precisely to probe the boundaries of OpenRoundup's utility, presented a more nuanced picture. These journalists weighed OpenRoundup against mature, habituated alternatives. They had already solved table consolidation using Pandas~\footnote{https://pandas.pydata.org} in Python or Dplyr~\footnote{https://dplyr.tidyverse.org} in R and the switching cost of adopting a second platform within an in-progress data pipeline outweighs the benefits of a graphical alternative. P1 noted that once data preparation has begun in a Python environment, there is no natural moment to pivot to a different tool.


While documenting data provenance is outside of our scope, reproducibility surfaced as a structural consideration. A journalist using OpenRoundup would need to maintain a separate data diary~\cite{malan-ask-2018} to document data provenance; GUI-driven wrangling has historically faced this criticism~\cite{sandve-ten-2013}. Code-based approaches produce both a transformed table and a record of the transformations performed, which allows editors to audit methodology and supports transparency for readers.  In practice, however, these scripts tend to be dataset-idiosyncratic: they encode assumptions about the schema and column value formats specific to the tables at hand, and rarely transfer to new datasets without significant revision. P3, who most closely matches the Scripter archetype but uses other GUI applications in exploratory contexts, acknowledged replicability as an issue but ``not a deal breaker.'' 

One further consideration surfaced in a problem area that is outside the scope of OpenRoundup's design. Value-level transformation must be completed before the tool can be used, because OpenRoundup operates at the schema level and focuses on interactive, graphical support for table consolidation rather than value transformation. 

\subsection{Results: Usability}
\label{sec:results-usability}

Participants were able to use OpenRoundup to combine their own data successfully during the tryout period. The balance between self-directed exploration and structural guardrails was explicitly valued: P3 contrasted it favorably with wizard interfaces, noting that ``it doesn't feel like you're going through a process that is already predetermined.'' Getting started was accessible; P3 described it as ``super easy to get started,'' and the drag-and-drop file upload was immediately understood. Following demonstration and the documentation, P2 found the learning curve manageable: ``once you kind of understand what the different parts are \ldots\ it's pretty, not too hard to learn.'' Performance on large files impressed P3---``50 megabyte file \ldots\ does the join \ldots\ pretty quickly.'' Row-count breakdown by match result was praised unprompted: P2 described it as ``clear and helpful.'' The column distribution cards were noted as useful for data exploration and spotting outliers, and multiple participants praised the tutorial quality as both helpful and motivational.

Deployment study participants raised several interface-level concerns. Some error messages failed to explain what went wrong or how to recover, so improving error and alert guidance would be productive future development. Visual complexity was a recurring concern, with P3 noting that many interface elements competed for attention simultaneously. Participants also raised lower-level issues around button placement, dialog design, default parameter values, and visual theming. Real workflows additionally surfaced operations, such as transposing wide-format as needed by P4, that require reshaping or aggregation before import. These operations are outside OpenRoundup's scope of interactive table consolidation, so they must be performed with external tools prior to use. We include a description of usability issues tangential to our inquiry in \SuppSecUsability.

\section{Discussion}
\label{sec:discussion}

OpenRoundup demonstrates that a no-code, browser-based system can support professional data journalists in ecologically valid table consolidation workflows, but its current design reflects deliberate tradeoffs. We discuss the system's constitutive limitations, then identify directions that would extend its reach: smarter guidance and large language model (LLM) integration, richer operations, and support for federated data sources.

\subsection{Limitations}
\label{sec:limitations}

OpenRoundup successfully supports professional data journalists in performing multi-table consolidation tasks on their own data, as demonstrated by the replication and deployment studies. The system's design reflects a deliberate set of tradeoffs, and several of its boundaries are constitutive of the approach rather than incidental.

The commitment to eager consolidation shifts preparation work upstream: users must resolve structural mismatches explicitly before merging. Users with source data that is not tidy~\cite{wickham-tidy-2014}, namely does not conform to Codd's 3rd normal form~\cite{codd-relational-1990}, found the negotiation burden heavier than a code-based approach would impose. 

Client-only, single-threaded JavaScript execution imposes a practical scalability ceiling from a computational point of view: very large files and computationally expensive operations such as cartesian-product joins can saturate the UI thread. DuckDB-WASM's 32-bit address space sets a hard upper bound of just under 2~GB across all loaded files combined. We set a limit of 100~MB per file following the convention of local file upload size limits with browsers, but that parameter could easily be increased.  
There is no hard limit on the number of tables. 
Visual scalability could be an issue in theory because the categorical color palette cycles after 10 operations, the point at which a workflow involves at least twenty tables. In practice, however, workflows remain below both computational and perceptual thresholds: all file sizes from the 17 workflows are under 100~MB, and all workflows required less than 10 operations.  

The system accepts only static, flat-file sources in ASCII-compatible formats as input; relational databases, hierarchically structured formats such as JSON and XML, live data APIs, and unstructured text are out of scope, though these exclusions reflect practical decisions about deployment complexity rather than fundamental architectural barriers. Alert coverage is scoped to the operation level, leaving data quality issues at the column or table level unsurfaced.

\subsection{Smarter guidance and alignment support} 
The deployment study revealed a consistent expectation gap: participants anticipated that the system would propose sensible initial parameters for operations rather than presenting blank controls. The necessary substrate for addressing this shortfall is a richer semantic column type model. OpenRoundup currently infers only two generic types, string and numeric, which is too coarse to support meaningful automated guidance during schema alignment or join key selection. Higher-level semantic types organized within a hierarchical type system that distinguish nominal identifiers, ordinal codes, calendar dates, and continuous measures would enable more targeted recommendations and more reliably surface the intended join key, reducing the configuration burden that participants found heaviest during the Pack operation.

This richer column metadata also constitutes a natural input surface for LLM-based assistance. AI coding tools are eroding the scripting fluency barrier that historically created an opening for GUI alternatives, but this dynamic creates an opportunity to embed LLM assistance within OpenRoundup rather than treating the two as competitors. OpenRoundup's bounded Saga API surface is well suited to LLM integration via the Model Context Protocol~\footnote{\url{https://modelcontextprotocol.io}}, which could support natural language operation construction, join key suggestion, and automated operation tree narration to address the reproducibility gap raised by participants. LLM-generated code is opaque by nature: a journalist relying on an AI coding assistant must trust that the produced script is correct, with limited ability to inspect or verify the logic prior to running it. This concern is significant since dirty data may lead to incorrect accusations, exposing news organizations to reputational and legal risk. OpenRoundup's visual, interactive operation tree addresses this issue directly: any operation that an LLM might propose can be rendered as a node in the tree, making the logic transparent and inspectable before the user commits to it. This visibility bridges the gulf of evaluation~\cite{norman-design-2013} in a way that autonomous code generation cannot. Our design principle remains the same: suggestions remain overridable, the basis for every recommendation is transparent to the user, and the analytical decisions remain the user's to make.

\subsection{Expanding the operation set} 
\label{secfuture-operations}

OpenRoundup's current operation set reflects the core table consolidation tasks most common in journalism practice. Pack is a binary operation with a fixed predicate vocabulary, and single-table transformations such as reshaping and aggregation must be performed upstream, outside the tool.

Two complementary extensions would meaningfully reduce these constraints. The first is richer multi-table join semantics: a variadic Pack, analogous to Stack's existing $n$-table semantics, that joins multiple table rows on a shared column key in a single schema configuration. This capability would address joining regional or diachronic datasets~\cite{kasica-dirty-2023} that share a key across many releases, reducing operation tree complexity without requiring a chain of binary Pack operations. Composite key support and approximate string-matching predicates would further reduce the preprocessing burden for the independently-published, messy datasets characteristic of journalism practice, where name columns often have spelling inconsistencies and encoding artifacts that foil exact matching. The second extension is single-table operations. Introducing reshape operations (pivot, unpivot) and multi-column aggregation as first-class nodes in the operation tree would reduce the upstream dependency on external tools, bringing more of the data preparation workflow within OpenRoundup's visual, inspectable model, but at the cost of additional operation tree complexity.

\subsection{Federated data sources} 

OpenRoundup currently requires users to download and import static files before consolidation can begin. The schema heterogeneity, naming inconsistencies, and differing column type conventions that characterize independently-maintained sources are precisely the problems OpenRoundup's alignment and consolidation model is designed to address, making federation a natural extension of the system's existing strengths.

Extending OpenRoundup to pull source tables directly from online data sources, such as government open data portals, collaborative spreadsheet platforms, remote columnar stores, or OLAP databases, would eliminate the need to download and re-import static files and position OpenRoundup as a \defn{data mediator}~\cite{garcia-database}, a software component that provides a unified virtual view over heterogeneous, independently-maintained sources. OpenRoundup's architecture supports this direction without changes to the Saga or State layers: the Wrapper contract already abstracts all persistent data store access behind a stable, five-operation interface, making each data source a pluggable backend. Federation would also strengthen provenance transparency: a consolidation that reads directly from an addressable source, rather than an imported snapshot, allows a journalist or editor to re-execute the workflow against updated data and verify that the output changes in a documented, predictable way. Any federated architecture must treat data privacy as a first-order constraint, preserving the user control that is central to journalism practice.

\section{Conclusion}
\label{sec:conclusion}

We present OpenRoundup, a browser-based system for multi-table data consolidation designed for data journalists without programming expertise. We contribute a conceptual framing for addressing the multi-table data integration problem in data journalism: \textit{eager table consolidation}, in which a composite table is incrementally assembled through a \textit{snowball approach} early so its unified schema guides all subsequent wrangling, 
and a declarative vocabulary for table consolidation consisting of two operations, Stack and Pack. 
We contribute OpenRoundup itself --- its interface design, architecture, and implementation --- as a system that instantiates this framing through a schema-first, values-on-demand paradigm. A replication study confirms its expressiveness through coverage of real journalist programming workflows. A deployment study with practitioners confirms utility in authentic reporting contexts. Together, these evaluations show that OpenRoundup addresses a recognized, persistent gap in data wrangling: handling a collection of tables, rather than treating a single table as the primary unit of work. It makes cross-source schema negotiation a first-class, accessible task for the reporters who need it most.

\section{Supplemental Material Index}

OpenRoundup is available as open source at \url{https://github.com/steve-kasica/roundup}, with a live demo at \url{https://cs.ubc.ca/group/infovis/openroundup}. Supplemental materials are available at \url{https://osf.io/mgu7c}, featuring further interface details; deployment study participant demographics, interview scripts, and usability issues surfaced; replication study material for the additional 16 journalist workflows including many OpenRoundup screenshots, and videos of three of the workflows.  

\bibliographystyle{IEEEtran_link}
\bibliography{strings, references}

\def\interBioSpace{-33pt} 
\vspace{\interBioSpace}
\begin{IEEEbiography}[{\includegraphics[width=1in,height=1.25in,clip,keepaspectratio]{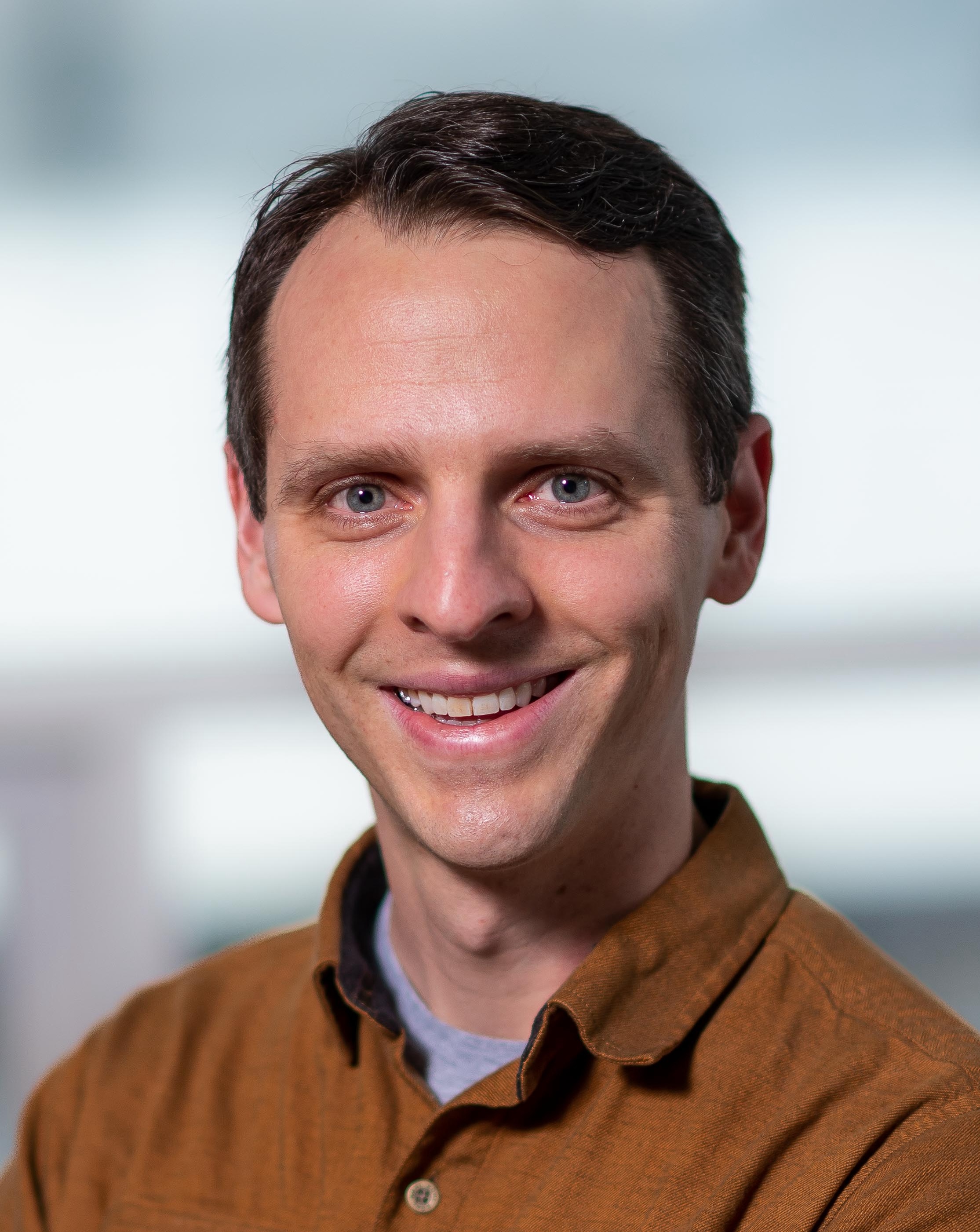}}]{Stephen Kasica} is a PhD candidate at the University of British Columbia, in Computer Science. His research focuses on using interactive visualization tools to support data journalists in their work, grounded in empirical studies of journalism practice. His prior research spans the design and evaluation of wrangling systems, observation studies of computational journalism, and interview-based investigations of data preparation in newsrooms.
\end{IEEEbiography}

\vspace{\interBioSpace}
\begin{IEEEbiography}[{\includegraphics[width=1in,height=1.25in,clip,keepaspectratio]{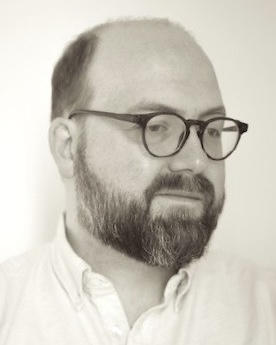}}]{Charles Berret} received the PhD degree from Columbia University in Communication Studies. He is a data journalist and digital media researcher who has taught at the University of British Columbia, Linköping University, and the School for Poetic Computation. He currently works at Enigma Technologies in New York.

\end{IEEEbiography}

\vspace{\interBioSpace}
\begin{IEEEbiography}[{\includegraphics[width=1in,height=1.25in,clip,keepaspectratio]{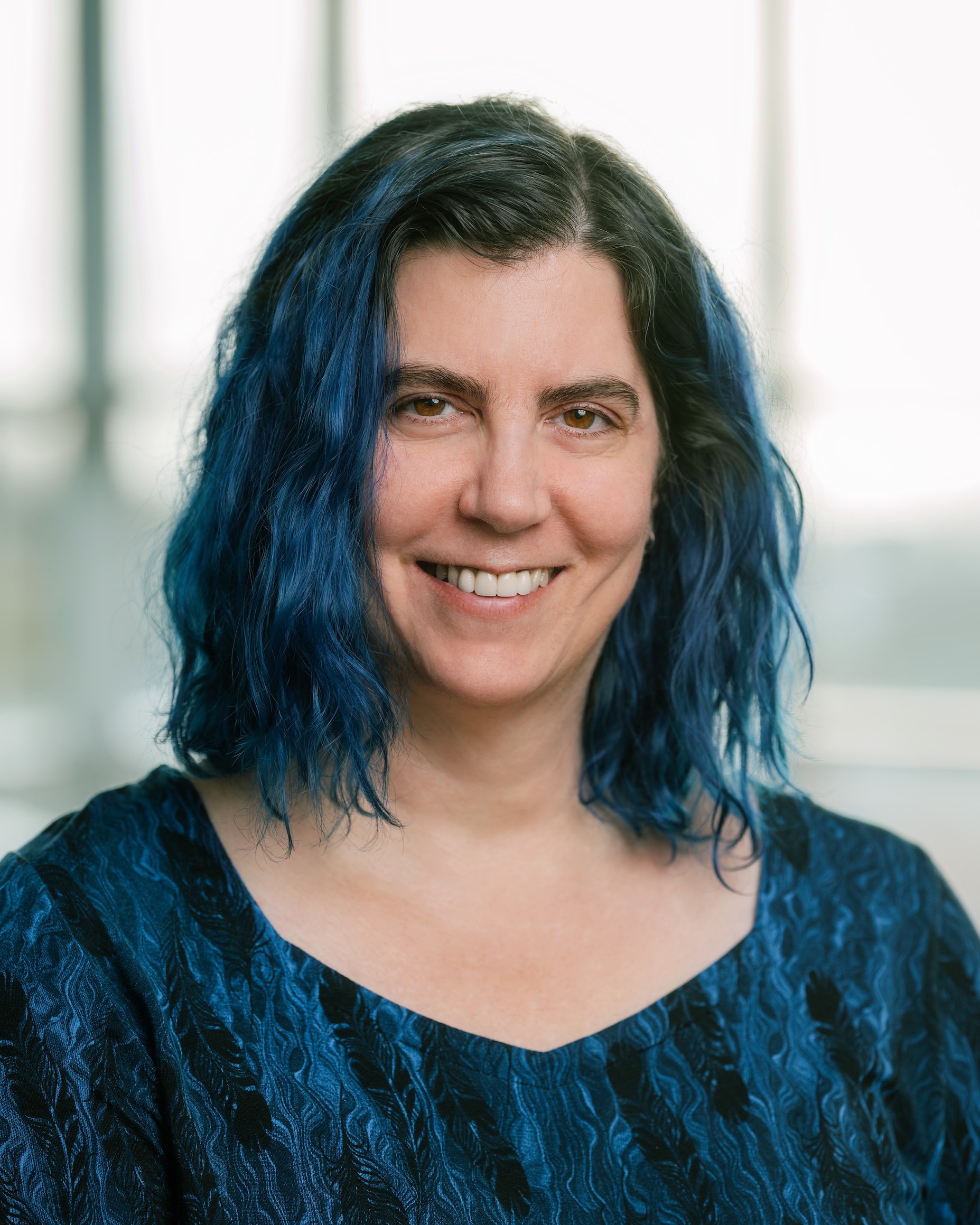}}]{Tamara Munzner} (IEEE Fellow) received the PhD degree from Stanford in Computer Science. She is currently a professor with the University of British Columbia. She has worked on visualization projects in a broad range of application domains from genomics to journalism. Her book Visualization Analysis and Design is heavily used worldwide, and she was the recipient of the IEEE VGTC Visualization Technical Achievement Award.

\end{IEEEbiography}

\vfill

\end{document}